\documentclass[tighten,twocolappendix,twocolumn]{aastex7}

\usepackage[T1]{fontenc}
\usepackage{amsmath}
\usepackage{xspace}
\usepackage{enumitem}
\usepackage{threeparttable}

\newcommand{\bhspin}{a_*}

\newcommand{\athenak}{{\tt AthenaK}\xspace}

\begin{document}

\shorttitle{Mass Transport and Mixing in Black Hole Accretion}
\shortauthors{Wong et al.}

\title{Mass Transport, Turbulent Mixing, and Inflow in Black Hole Accretion}

\author[0000-0001-6952-2147]{George N.~Wong}
\email{gnwong@ias.edu}
\correspondingauthor{George N.~Wong}
\affiliation{School of Natural Sciences, Institute for Advanced Study, 1 Einstein Drive, Princeton, NJ 08540, USA}
\affiliation{Princeton Gravity Initiative, Princeton University, Princeton, New Jersey 08544, USA}

\author[0000-0003-2342-6728]{Lia~Medeiros}
\email{lia2@uwm.edu}
\affiliation{Center for Gravitation, Cosmology and Astrophysics, Department of Physics, \\University of Wisconsin–Milwaukee, P.O. Box 413, Milwaukee, WI 53201, USA}
\affiliation{Department of Astrophysical Sciences, Peyton Hall, Princeton University, Princeton, NJ, 08544, USA}

\author[0000-0001-5603-1832]{James~M.~Stone}
\email{jmstone@ias.edu}
\affiliation{School of Natural Sciences, Institute for Advanced Study, 1 Einstein Drive, Princeton, NJ 08540, USA}

\begin{abstract}
We investigate mass transport, mixing, and disk evolution in non-radiative black hole accretion flows using Lagrangian tracer particles embedded in general relativistic magnetohydrodynamics simulations. Our simulation suite spans magnetically arrested disk (MAD) and standard and normal evolution (SANE) states across a range of black hole spins. By tracking tracer trajectories, we directly measure both advective inflow and stochastic spreading of fluid elements. The tracer distributions are well described by a combination of coherent inward drift and Gaussian-like broadening, consistent with an advection-diffusion picture. MADs exhibit systematically faster inflow than SANEs, with retrograde flows showing the most rapid infall; the innermost stable circular orbit leaves little imprint in MADs but remains more visible in SANEs. Turbulent fluctuations drive strong radial dispersion in all cases, with a superdiffusive scaling of $\sigma \propto t^{\,0.95}$ in MADs and $\sigma \propto t^{\,0.75}$ in SANEs for high-spin prograde disks. Mixing times decrease toward the event horizon and are consistently shorter in MADs and retrograde configurations. Tracers also reveal how accretion sources shift over time: turbulence draws inflow from a broad range of initial radii, with rapid torus depletion in MADs driving the mean source radius outward as $r_{\rm cyl} \propto t^{\,2/3}$, while SANEs evolve more gradually with $r_{\rm cyl} \propto t^{\,1/2}$. We show that the finite mass of the initial torus has a strong influence on late-time behavior, especially in MADs, where imprints of differently sized initial conditions may be accessible as early as $t \approx 10^4\ GM/c^3$.
\end{abstract}

\keywords{accretion (14) -- magnetohydrodynamics (1964) -- astrophysical black holes (98) -- magnetohydrodynamical simulations (1966)}

\section{Introduction}

Black hole accretion powers some of the most energetic astrophysical phenomena in the universe. From stellar mass black hole binaries to active galactic nuclei, the gravitational energy released by inflowing gas drives intense radiation, outflows, and in some cases, powerful relativistic jets. The physics governing these systems is complex, involving a delicate interplay of mass, energy, angular momentum, and magnetic fluxes. The accretion process itself comprises a competition between coherent inflow and stochastic transport. Magnetic turbulence and viscous effects drive angular momentum exchange, while shocks, interchange motions, and large-scale magnetic stresses redistribute mass, energy, and magnetic flux. These processes set inflow times, control how quickly plasma from different parts of the disk can communicate and mix, and ultimately shape the variability and outflows probed by observations. 

The basic mechanisms underlying accretion are understood. As plasma spirals inward toward the central object, its gravitational potential energy is converted into kinetic and thermal energy. For accretion to proceed through nearly circular orbits, this energy must be removed, either by radiating it away or by advecting it across the event horizon. At the same time, matter must rid itself of angular momentum: in a nearly Keplerian disk, gas at larger radii carries more angular momentum, so material must shed angular momentum in order to move inward. Dissipation occurs through processes such as viscous shear, turbulence, shocks, and magnetic reconnection, producing the broad band emission that characterizes accreting black holes. A fundamental requirement, then, is the outward transport of angular momentum to allow mass to spiral inward while releasing gravitational energy.

The earliest analytic models addressed this by parameterizing angular momentum transport as an effective viscosity. The $\alpha$-disk model of \cite{shakura_1973_disks}, and its relativistic extension by \cite{novikov_1973_diskmodel}, captured the essential energetics of thin, radiatively efficient accretion disks, where local dissipation is assumed to be rapidly radiated away. These models remain enormously influential, but they idealize transport as a local shear stress, and are agnostic about the physical origin of angular momentum exchange. Subsequent work established that magnetic fields likely play a central role: the magnetorotational instability efficiently taps the free energy of differential rotation to drive turbulence and outward angular momentum transport, a result now confirmed in local and global simulations \citep{velikhov_1959_mri,chandrasekhar_1960_mri,balbus_1991_mri,hawley_1995_localmri,stone_1996_stratified,hawley_2000_globaltori}.

In radiatively inefficient accretion flows (RIAFs), as is the nature of many low-accretion rate systems, the gas cannot cool efficiently and the resultant retained thermal energy provides a pressure support that leads to a geometrically thick disk. In the subclass of advection-dominated accretion flows, most of this energy remains bound to the plasma and is ultimately carried through the event horizon \citep{narayan_1995_ADAF}. In these hot, low-luminosity systems, angular momentum transport is thought to be governed by magnetic turbulence, magnetic braking, and large-scale interchange instabilities, though the relative importance and interplay of these processes remain open questions.
RIAFs are typically modeled using general relativistic magnetohydrodynamics (GRMHD) simulations, which evolve the coupled plasma and magnetic field in three dimensions. These simulations self-consistently produce turbulent, time-dependent flows and enable detailed predictions of disk structure, winds, and relativistic jets. As a result, GRMHD modeling has become the standard tool for studying RIAFs (e.g., \citealt{gammie_2003_harm,mignone_2007_PLUTO,delzanna_2007_ECHOEulerianConservative,narayan_2012_sane,sadowski_2013_koral,white_2016_athenapp,porth_2017_BHAC,liska_2018_tilted,cipolletta_2021_spritz,shankar_2023_gramx,stone_2024_athenak}).

Despite decades of study, many fundamental questions about black hole accretion remain unresolved. While it is broadly understood that accreting gas loses angular momentum while releasing gravitational energy, the quantitative details of the turbulence, magnetic fields, and dissipative heating remain uncertain. These uncertainties have direct consequences for interpreting high-resolution observations of black hole systems, from the properties of the variability and the composition of the emitting plasma to the structure of horizon-scale images. Understanding how matter and magnetic flux circulate through the disk is therefore central to explaining a wide range of observed astrophysical phenomena.

Many of the outstanding questions concern the transport and mixing of mass within the disk. Can the plunging region inside the innermost stable circular orbit (ISCO) be distinguished from the rest of the inflow in geometrically thick disks? How efficiently does turbulence redistribute material, and on what timescales do fluid elements mix or migrate inward? What physical processes control these rates, and how do the rates differ across magnetic flux configurations and black hole spins? Is the notion of an ``inflow time’’ reasonable in such turbulent environments, and how do finite-mass tori (the usual initial condition in GRMHD simulations) bias the apparent draining rates, source radii, and scaling laws we measure? 

These questions matter not only for accretion theory but also for connecting simulations to observations. The degree of turbulent mixing can regulate thermal equilibration, chemical composition transport, and the coherence of inflow streams, all of which have potential observational signatures; mixing and inflow efficiency also affect jet entrainment profiles, which help determine the source of matter in relativistic outflows \citep[e.g.][]{yuan_2003_nonthermal,mckinney_2006_jets,ressler_2015_electronthermo,sadowski_2017_twotemp,wong_2021_jetdisk}. In more extreme contexts, such as nuclear-dominated disks formed in compact object disruptions, mixing can control whether nuclear burning ignites and fundamentally alters the flow dynamics \citep{fernandez_2013_delayed,fernandez_2015_outflowsnsmergers,margalit_2016_nuclearburning}.

Yet the question of mass mixing in accretion flows has not been studied in great depth. Most prior work has focused on global properties such as angular momentum transport or large-scale inflow/outflow rates, often within the context of shearing box simulations that probe local turbulence and magnetorotational instability in detail but do not capture the full global dynamics (see \citealt{carballido_2005_localdiffusion,turner_2006_mixing,kapyla_2009_turbdiffusion,tominaga_2019_dustdiffusion} for various contexts). While traditional Eulerian diagnostics are useful for understanding instantaneous fluxes of mass and energy, they do not capture the histories of individual fluid parcels. To overcome this limitation, we adopt a complementary Lagrangian perspective by embedding passive tracer particles into global GRMHD simulations of radiatively inefficient accretion flows. These tracers follow the motion of fluid elements throughout the domain, allowing us to directly measure advection and mixing over time. With this approach, we can quantify the spatial and temporal origins of disk material, trace its acceleration history, and characterize how gas is transported, mixed, and ultimately advected through the black hole event horizon or lost in outflows.

Our simulation suite systematically explores radiatively inefficient accretion flows in both the magnetically arrested disk (MADE) and standard and normal evolution (SANE) configurations. We span black hole spin values from $\bhspin = 0$ to $0.9$ for both prograde and retrograde disk orientations in order to capture the full range of dynamical behaviors. To ensure robustness, we conduct a dense survey of models at standard resolution, complemented by a targeted set of higher-resolution simulations to test convergence of our transport diagnostics. This design allows us not only to map out trends across magnetic state and spin but also to establish confidence in the reliability of our tracer-based mixing and inflow measurements.

The remainder of this paper is organized as follows. In Section~\ref{sec:methods}, we describe the numerical methods for the fluid and tracer particles and then summarize the details of our simulation suite. We then present our results in two parts. First, in Section~\ref{sec:advection_mixing}, we analyze the local properties of mass transport and redistribution within the disk and report on radial advection and mixing. Then, in Section~\ref{sec:disk_evolution}, we consider the global evolution of the system, measuring infall times, draining of the initial torus, and the dependence of these processes on black hole spin and magnetization state. Finally, in Section~\ref{sec:discussion}, we discuss the implications and limitations of our findings and outline directions for future work.

\section{Numerical Methods}
\label{sec:methods}

We perform a suite of GRMHD simulations of radiatively inefficient accretion flows onto spinning supermassive black holes in the Kerr geometry using the \athenak code \citep{stone_2024_athenak}. Our setup includes 48 primary simulations spanning a range of black hole spins and magnetic field configurations across two different resolutions. To enable Lagrangian diagnostics of fluid motion, we augment the code with a Lagrangian Monte Carlo tracer particle scheme following the method described in \citet{genel_2013_lagrangianmc}. In this section, we detail the numerical implementation of the fluid and particle modules and describe the details of our simulation library. More detail about the simulation methodology and toolkit can be found in \citet{wong_2022_patoka,stone_2024_athenak}.

\subsection{GRMHD}

We evolve the accretion flow by solving the equations of ideal general relativistic magnetohydrodynamics (GRMHD) using the GPU-accelerated code \athenak. All simulations are performed in ingoing horizon-penetrating Kerr-Schild coordinates, and the spacetime is fixed according to the Kerr metric with spin parameter $\bhspin = J c / GM^2$. The equations are formulated as a set of hyperbolic conservation laws written in a general coordinate basis:
\begin{align}
\partial_t \left( \sqrt{-g} \rho u^t \right) &= -\partial_i \left( \sqrt{-g} \rho u^i \right), \label{eqn:massConservation}\\
    \partial_t \left( \sqrt{-g} {T^t}_{\nu} \right) &= - \partial_i \left( \sqrt{-g} {T^i}_{\nu} \right) + \sqrt{-g} {T^{\kappa}}_{\lambda} {\Gamma^{\lambda}}_{\nu\kappa},  \\
\partial_t \left( \sqrt{-g} B^i \right) &= - \partial_j \left[ \sqrt{-g} \left( b^j u^i - b^i u^j \right) \right], \label{eqn:fluxConservation}
\end{align}
along with the constraint
\begin{align}
\partial_i \left( \sqrt{-g} B^i \right) &= 0. \label{eqn:monopoleConstraint} 
\end{align}
Here, $\rho$ is the rest-mass density, $u^\mu$ is the fluid four-velocity, and $b^\mu$ is the magnetic four-vector for the field in the comoving frame. The determinant of the metric is $g$, and ${\Gamma^\lambda}_{\mu\nu}$ are the Christoffel symbols of the background spacetime. The total stress-energy tensor is given by
\begin{align}
{T^{\mu\nu}} &= \left( \rho + u + P + b^{\lambda}b_{\lambda}\right)u^{\mu}u^{\nu} \nonumber  \\
&\qquad \quad + \left(P + \frac{b^{\lambda}b_{\lambda}}{2} \right){g^{\mu}}^{\nu} - b^{\mu}b^{\nu},
\label{eqn:mhdTensor}
\end{align}
where $u$ is the internal energy and the fluid pressure is obtained via the ideal gas law $P = (\hat{\gamma} - 1) u$ with adiabatic index $\hat{\gamma}$. In our simulations, we set $\hat{\gamma} = 13/9$, which is appropriate for a plasma with the same ion and electron temperatures but relativistic electrons and non-relativistic ions (but see \citealt{gammie_2025_adiabaticindex} for more detail about this choice). Our simulations are scale-free in that they are invariant under rescalings of both length and mass density. This assumption is appropriate for the low-accretion-rate RIAFs and ADAFs we target and that are thought to characterize a significant fraction of nearby galaxies \citep{ho_2008_llagn}.

\athenak solves the GRMHD equations on a logically Cartesian mesh in Cartesian Kerr–Schild (CKS) coordinates with static mesh refinement, concentrating resolution at small radii near the black hole where dynamical timescales are shortest and the flow is most variable. In practice, the lowest-resolution (coarsest) refinement grid covers a cube centered on the black hole, and each successive refinement level spans a concentric cube with side lengths half those of its parent.  
See Appendix~\ref{sec:appendix_grid_comparison} for a comparison between the CKS grid used by \athenak and the exponential modified spherical Kerr-Schild grids used by other codes, such as Kharma \citep{prather_2024_kharma}. 

The magnetohydrodynamics sector is evolved using the constrained transport scheme of \citet{evans_1988_ct} with upwinded electric fields computed following \citet{gardiner_2005_ct,gardiner_2008_ct}. \athenak also employs the first-order flux correction (FOFC) fallback described by \citet{lemaster_2009_fofc}, which is only invoked in rare instances where zone-averaged density or pressure would otherwise become negative. The FOFC scheme reduces the reliance on numerical floors and thereby improves the conservative accuracy of the solution at the cost of increased diffusivity. We adopt a density floor of $\rho_{\rm floor} = 10^{-8}$ and a pressure floor of $p_{\rm floor} = \tfrac{1}{3} \times 10^{-10}$, which adjusts the internal energy of the fluid. For flux computations, we use the HLLE Riemann solver in combination with fourth-order piecewise parabolic reconstruction.

\subsection{Particles}

To measure mass transport, mixing, and the histories of accreting material, we embed passive tracer particles in each GRMHD simulation using a newly implemented particle module. The particles are evolved alongside the fluid but exert no dynamical influence on it, so they act as Lagrangian tracers of the underlying flow. Each tracer is assigned a persistent unique identifier, which enables us to reconstruct complete particle trajectories and histories throughout the simulation. Because we can recover and trace the mass flow throughout the temporal evolution of the simulation, this approach allows us to directly identify the origin of accreted matter and is robust against artifacts introduced by density or energy floor prescriptions in the fluid evolution.

We employ a Lagrangian Monte Carlo method for particle transport, following the formalism of \citet{genel_2013_lagrangianmc}. Rather than updating particle positions by interpolating the local fluid velocity, we instead determine particle motion probabilistically from the conserved mass fluxes computed during each GRMHD update step. At each timestep, the outgoing mass fluxes through a zone's interfaces are normalized by the total conserved mass in the zone. Each particle within the zone then draws a random number to select its destination, either remaining in the same zone or moving to a neighboring one. This ensures that the particle population follows the mass flow, avoiding systematic biases in mass distribution and power spectra that can arise in velocity-interpolation schemes (see \citealt{genel_2013_lagrangianmc} for more detail).

The particle update step is executed every fluid timestep and is fully parallelized across GPUs. Communication of particle data between MPI ranks uses a gather–scatter operation to minimize message passing. Our particle module also supports mesh refinement following the same rules as for regular transport: When a particle moves into a more refined block, it is probabilistically placed in one of the child cells according to the incident flux ratios; when moving to a coarser block, it is assigned to the corresponding parent cell.

We initialize particles by distributing them throughout the initial torus with probability proportional to the coordinate mass density $\rho u^t \sqrt{-g}$, where $\rho$ is the rest-mass density, $u^t$ is the time component of the four-velocity, and $g$ is the determinant of the covariant metric. This yields an approximately constant number of particles per unit coordinate mass. Particle data are recorded at regular output intervals; we also store the minimum radius attained by each particle over its lifetime as well as the precise time when it reached that radius.

The Lagrangian Monte Carlo scheme introduces some artificial diffusivity into particle trajectories due to its stochastic update rule. This manifests as a random-walk component in individual histories, particularly over long timescales, and cannot be mitigated by simply increasing the particle count. Nonetheless, the numerical diffusivity is lower than the fluid diffusivity, and the ensemble-averaged quantities that form the basis for the diagnostics in this paper, such as advection velocities, spread rates, and inflow/outflow fractions, are recovered accurately.

The tracer particle module in \athenak is designed for minimal overhead in large-scale GRMHD calculations. In the highest-resolution simulations, the combined particle push, communication, and I/O steps account for only $\sim5\%$ of the total wall-clock time per fluid timestep, with the cost scaling linearly with the total number of particles. On NVIDIA A100 GPUs, for our high resolution simulations with $10^7$ particles spread over $8$ GPUs across two nodes, the particle update rate reaches approximately $2.2 \times 10^7$ particle updates per second, while the fluid solver sustains roughly $3.3 \times 10^8$ zone updates per second. These numbers represent the total performance of the full code, including both fluid and particle updates. When measured in isolation, the update rates for each component are correspondingly higher. Appendix~\ref{sec:appendix_grid_comparison} provides more detail on the runtime and number of updates required for different simulations.

\subsection{Simulations overview}

\begin{figure*}
\begin{centering}
\includegraphics[width=0.95\linewidth]{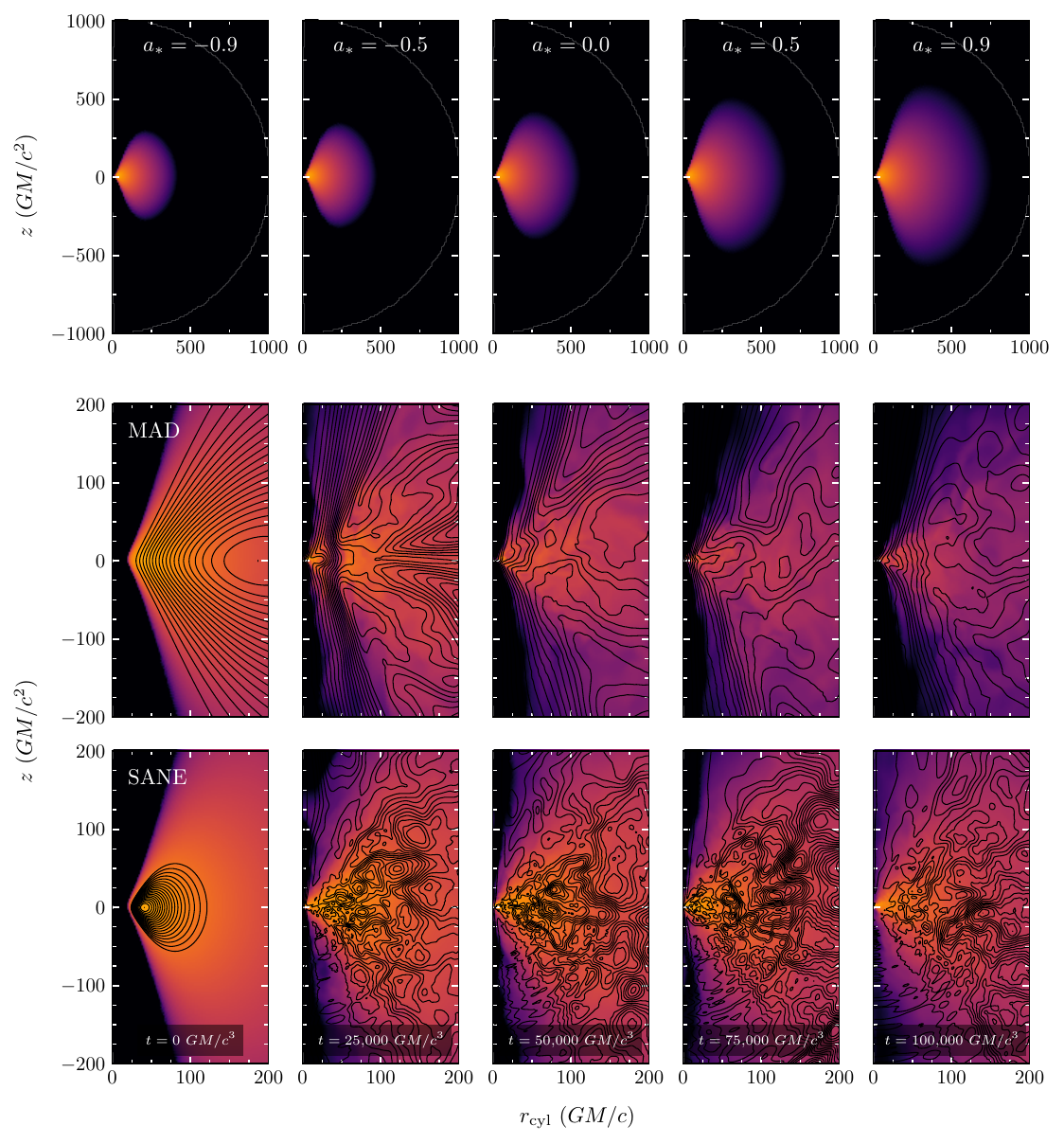}
\caption{
Rest-mass plasma density across simulations and times. Colors span logarithmically over five orders of magnitude and are normalized across the entire figure rather than per-panel. Each panel shows a vertical slice through the $r_{\rm cyl}-z$ plane, evaluated in Kerr-Schild coordinates. Top: initial condition for the usual Fishbone-Moncrief torus with $r_{\rm in} = 20$ and $r_{\rm max} = 41$ across five black hole spins $-0.9, -0.5, 0, 0.5, 0.9$, where negative numbers indicate that the plasma rotates opposite the spin of the central black hole. Notice that the size of the initial torus depends on the angular momentum of the central black hole, but the same fluid initial condition is used for both MAD and SANE configurations. Middle/Bottom: Zoomed-in representation of the time evolution of plasma with lines showing the axisymmetrized magnetic field overplotted in black for $\bhspin = 0.9$ MAD (middle) and SANE (bottom) simulations. The plasma is evaluated on an azimuthal slice. The left-most panels show the initial condition at $t = 0\ GM/c^3$, and the subsequent columns show the simulation at times $t = 25,\!000,\, 50,\!000,\, 75,\!000,$ and $100,\!000\ GM/c^3$.
}
\label{fig:grmhd_overview}
\end{centering}
\end{figure*}

\begin{figure}
\begin{centering}
\includegraphics[width=\linewidth]{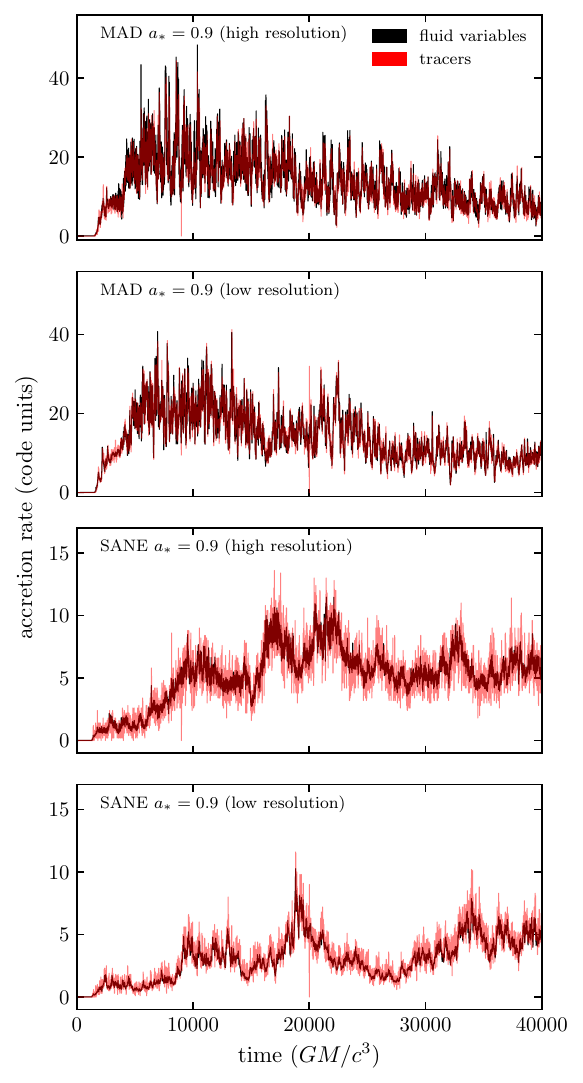}
\caption{
Comparison of accretion rates from GRMHD fluid variables (dark) and from tracer particles crossing the event horizon (light red) for representative high-spin $\bhspin = 0.9$ MAD (top) and SANE (bottom) simulations at both $16/8$ and $8/8$ resolutions (see main text). For the higher-resolution runs, the tracer-based rate is sampled twice as frequently, while the fluid accretion rate is a summary statistic reported with the same output cadence across simulations. The close agreement between tracer- and fluid-based measurements confirms that the tracer population accurately captures the mass inflow and that the initial particle distribution was appropriately chosen to follow the fluid dynamics.
}
\label{fig:accretion_rates_comparison}
\end{centering}
\end{figure}

We use \athenak to perform a suite of 48 three-dimensional GRMHD simulations of radiatively inefficient accretion flows, spanning a wide range of black hole spins and magnetic field configurations. The computational domain extends over a cube of side length $2048\ GM/c^2$, reaching to $\pm 1024\ GM/c^2$ in each CKS coordinate direction. Outflow boundary conditions are imposed at the domain edges, while the central black hole is treated with an excision boundary \citep{olivares_2019_bhac,ressler_2021_sphericalaccretion,stone_2024_athenak}. To assess numerical convergence, we adopt two grid resolutions defined within the innermost $\pm 8\ GM/c^2$ region: a standard resolution of 8 cells per $GM/c^2$ and a higher-resolution variant with 16 cells per $GM/c^2$. All simulations are seeded with $10^7$ tracer particles distributed proportionally to the coordinate mass density. We represent the black hole angular momentum by the dimensionless spin parameter $\bhspin \equiv Jc/GM^2$, with $\left|\bhspin\right|\leq 1$ (using units $G=c=1$). Negative values of $\bhspin$ correspond to retrograde configurations in which the black hole spin is anti-aligned with the angular momentum of the inflowing gas at large radii.

The simulation initial condition also depends on the net magnetic flux threading the black hole horizon. In ideal MHD, accreting plasma drags magnetic field lines inward and magnetic flux accumulates until the magnetic pressure near the event horizon balances the ram pressure of the infalling gas. This state, known as the magnetically arrested disk (MAD; \citealt{bisnovatyi_1974_madstar,igumenshchev_2003_mad,narayan_2003_mad}), is characterized by intermittent accretion mediated by narrow streams of accreting plasma. The complementary regime is known as the standard and normal evolution (SANE) state, where the magnetic flux on the horizon remains comparatively low and the flow evolves more smoothly with turbulent but quasi-steady inflow \citep{narayan_2012_sane,sadowski_2013_sane}. MAD-like flows are particularly well motivated by recent Event Horizon Telescope results for M87$^*$ and Sgr A$^*$, which favor scenarios with strong horizon-threading fields \citep{eht_m87_5,eht_m87_8,eht_m87_9,eht_sgra_5,eht_sgra_8}. In order to achieve the MAD or SANE state, we seed our simulations with an axisymmetric vector potential defined by either
{\small
\begin{align}
A_{\phi;\, \rm SANE} &= {\rm max} \left[ \dfrac{\rho}{\rho_{\rm max}} - 0.2, 0 \right] \\
A_{\phi;\, \rm MAD} &= {\rm max} \left[ \dfrac{\rho}{\rho_{\rm max}} \left( \dfrac{r}{r_0} \sin\theta\right)^3 e^{-r/400} - 0.2, 0 \right].
\end{align}}
In total, our simulation suite spans both MAD and SANE configurations across $\bhspin$ values from $-0.9$ to $+0.9$, encompassing both prograde and retrograde orientations and includes
\begin{enumerate}[leftmargin=2em]
\item 38 standard-resolution simulations covering spins $a_* = -0.9$ to $+0.9$ in increments of $0.1$ for both MAD and SANE configurations, and
\item 10 high-resolution simulations for both MAD and SANE configurations at five representative spins $a_* = -0.9$, $-0.5$, $0$, $+0.5$, and $+0.9$.
\end{enumerate}

We initialize our simulations using the hydrostatic equilibrium torus solution of \citet{fishbone_1976_torus} with the inner edge located at $r_{\rm in} = 20\ GM/c^2$ and the pressure maximum at $r_{\rm max} = 41\ GM/c^2$. We evolve each simulation for at least $40,\!000\ GM/c^3$ to allow the system to settle into a quasi-steady, turbulent state and to enable us to study how the initial torus drains. To probe long-term disk behavior, we run the highest-spin prograde MAD and SANE standard simulations to $100{,}000\ GM/c^3$. A summary of the simulation suite, including varying parameters and output cadence, is given in Table~\ref{table:sim_summary}. For context, fluid snapshots are $3.6 {\rm GB}$ in the high-resolution runs and $457\ {\rm MB}$ in the standard-resolution runs, and particle snapshots are each $344 {\rm MB}$ in size.

Figure~\ref{fig:grmhd_overview} provides a guide to both the structure of the initial torus and the subsequent evolution of the accretion flow across our simulation suite. The panels show azimuthal slices of the rest-frame mass density in logarithmic scale over five orders of magnitude. The initial torus size depends on the black hole spin, with prograde configurations having larger initial disks. The bottom two rows show the evolution of the tori and overplot the axisymmetrized magnetic field geometry. 

\begin{deluxetable*}{ cccccc }
\tablecaption{Summary of simulations in this work} \label{table:sim_summary}
\tablehead{ 
\colhead{resolution} &
\colhead{black hole spin $\bhspin$} & 
\colhead{duration} &
\colhead{fluid cadence} &
\colhead{particle cadence} &
\colhead{notes} 
}
\startdata
$8/8$ & $0.9$ & 100,000 $GM/c^3$ & 50 $GM/c^3$ & 10 $GM/c^3$ & MAD and SANE (2 sims) \\
$8/8$ & $-0.9$ to $0.8$ (step $0.1$) & 40,000 $GM/c^3$ & 50 $GM/c^3$ & 10 $GM/c^3$ & MAD and SANE (36 sims) \\
$16/8$ & $-0.9, -0.5, 0, 0.5, 0.9$ & 40,000 $GM/c^3$ & 20 $GM/c^3$ & 5 $GM/c^3$ & MAD and SANE (10 sims) \\
\enddata
\tablecomments{Overview of simulation parameters for the suite considered in this paper. Simulations are either standard (8/8) or high resolution (16/8), and run for a default $40,000\ GM/c^3$ duration or for an extra ``long-duration'' time range until $100,000\ GM/c^3$. Each simulation tracks $10^7$ individual tracer particles and saves them at the rate specified in the table. We include a full grid over spin $-0.9 \le \bhspin \le 0.9$ in steps of 0.1 at lower resolution and a more limited sampling at higher resolution. All simulations are run for both the MAD and SANE magnetic field configuration, with an adiabatic index of $\hat{\gamma} = 13/9$, initialized from a \citet{fishbone_1976_torus} torus with $r_{\rm in} = 20\ GM/c^2$ and pressure maximum at $r_{\rm max} = 41\ GM/c^2$.}
\end{deluxetable*}

As an initial validation of our tracer particle module and to demonstrate consistency across grid resolutions, in Figure~\ref{fig:accretion_rates_comparison}, we plot the mass accretion rate through the event horizon
\begin{align}
\dot{M} = \int\limits_0^{2\pi}\int\limits_0^\pi \rho u^r \sqrt{-g} \ \mathrm{d} \theta \, \mathrm{d} \phi
\label{eqn:mdot}
\end{align}
as computed directly from the GRMHD variables or from the tracer particles. In the latter case, we estimate the accretion rate by counting the number of tracers that cross the event horizon within short time windows. Note that the tracer output cadence is twice as frequent in the high-resolution simulations as in the standard-resolution ones. At both resolutions, the two measurements are consistent, confirming that the initial tracer distribution and subsequent transport are handled correctly. MAD simulations show higher mean accretion rates in code units, with stronger short-timescale variability, while SANE disks accrete more steadily. In both regimes, the accretion rate also exhibits a gradual secular trend that infers disk draining.

\begin{figure*}
\begin{centering}
\includegraphics[width=\textwidth]{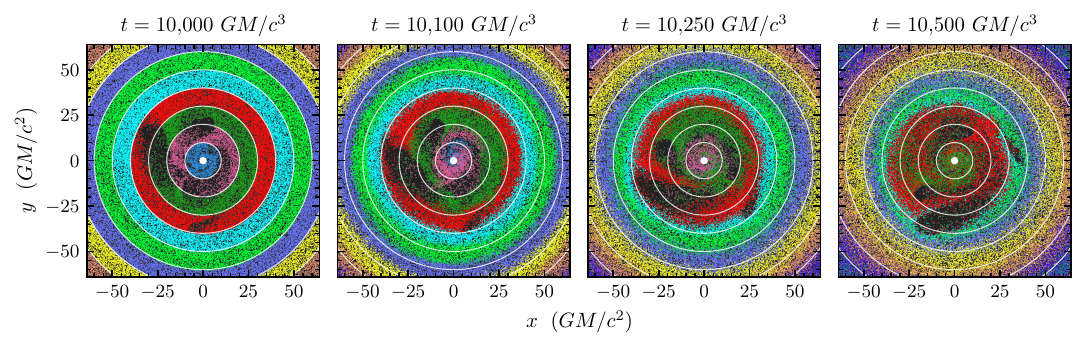}
\caption{
Evolution of tracer particles over time in the high-spin prograde $\bhspin = 0.9$ MAD simulation. Each panel shows the positions of tracer particles within a narrow wedge about the midplane ($|z/r| \leq 0.1$), colored according to their cylindrical radius at the initial time. The white filled circle at the origin marks the black hole event horizon. White rings overlaid on each panel indicate the initial radial boundaries between colored bins. As time progresses, the tracer population becomes increasingly mixed, demonstrating the radial redistribution of material as it is accreted toward the black hole.
}
\label{fig:tracer_spread_splash}
\end{centering}
\end{figure*}

\section{Advection and Mixing in the Disk}
\label{sec:advection_mixing}

We begin by analyzing the local properties of mass transport in the disk by exploring tracer particle histories. The details of the transport depend on both the magnetization state of the flow and on black hole parameters such as spin. We separate the motion into two principal components: a coherent inward advection that characterizes the inflow and a stochastic spreading that mixes neighboring fluid parcels.

\subsection{Radial mass transport in the disk}

To characterize the dynamics of the accreting plasma, we examine how fluid parcels passing through different radii migrate and spread over time. Tracer particles allow us to follow these Lagrangian displacements directly, which provides a direct way to distinguish between two primary modes of transport: coherent inward advection and stochastic redistribution driven by turbulence and other non-stationary fluctuations. Figure~\ref{fig:tracer_spread_splash} illustrates this process in the high-spin prograde ($\bhspin=0.9$) MAD simulation. Particles that lie within the wedge $h/r \le 0.1$ at each time are shown, color-coded by their cylindrical radius at $t=10,\!000\ GM/c^3$. As the simulation evolves, material from a wide range of radii is steadily drawn toward the black hole while turbulent motions cause the colors to spread and interleave, signaling strong radial mixing.

\begin{figure*}
\begin{centering}
\includegraphics[width=0.95\textwidth]{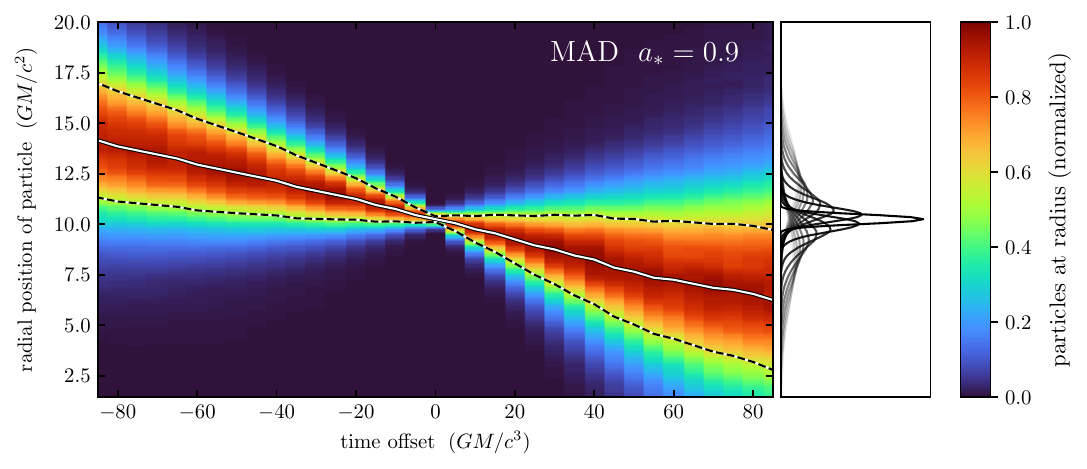}
\includegraphics[width=\linewidth]{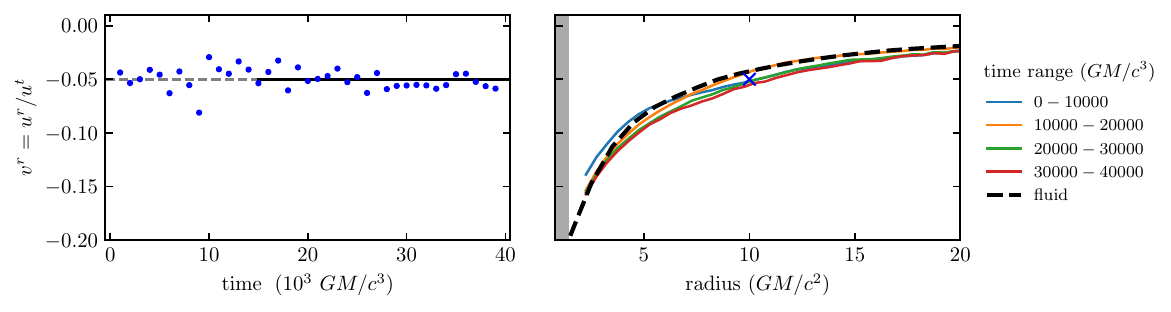}
\caption{
Mass advection and radial spreading in the high-spin ($\bhspin = 0.9$) high-resolution MAD simulation. As an illustrative case, the top panels track all particles crossing the radial shell $10 \le r < 10.5\ GM/c^2$. Each trajectory is time-shifted so that its passage through the shell is aligned, and the resulting histogram of radial positions as a function of time offset is shown in the top-left panel. The top-right panel displays the corresponding (unnormalized) radial distribution as it evolves with time.
The ensemble of trajectories demonstrates that material in the MAD disk is subject to both rapid inward advection and strong radial mixing. The mean particle trajectory (solid white line, top panel) indicates a systematic inward drift consistent with the net advection rate, while the growing spread (dashed white lines) reflects the rate of radial mixing.
This spreading occurs on timescales much shorter than the global accretion time, suggesting strong mixing in the flow.
The inferred advection velocity $v^r = u^r/u^t$ (bottom left) is consistent over $t = 15,\!000-40,\!000\ GM/c^3$ (black line) and matches the radial velocity profile measured directly from the GRMHD fluid variables (bottom right, dashed black). Agreement between tracer-based and fluid-based measurements confirms that the tracer population reliably follows the mass flow. Advection rates cannot be measured at small radii because particles are no longer evolved after crossing the event horizon.
}
\label{fig:MAD_0p9_advection}
\end{centering}
\end{figure*}

\begin{figure*}
\begin{centering}
\includegraphics[width=0.95\textwidth]{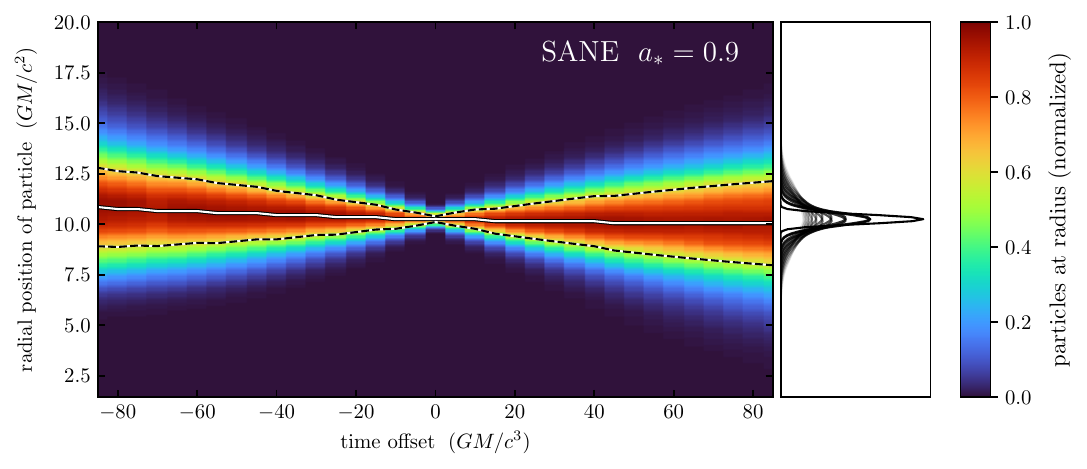}
\includegraphics[width=\linewidth]{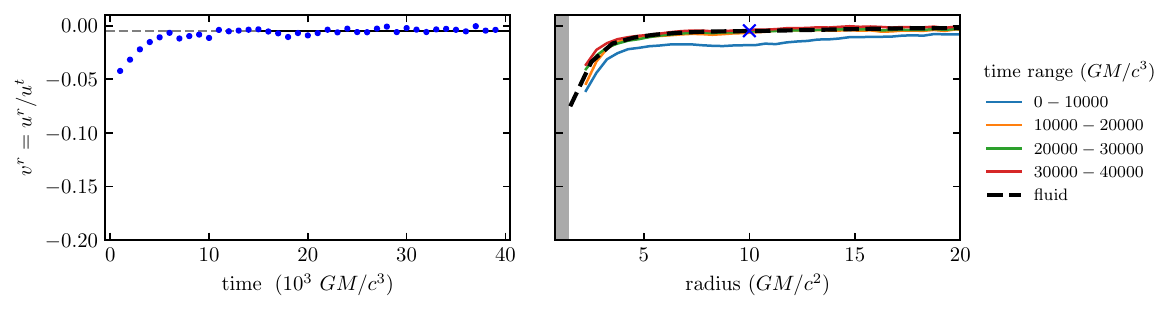}
\caption{
Same as Figure~\ref{fig:MAD_0p9_advection} but for the high-spin $\bhspin = 0.9$ high-spin SANE simulation. As can be seen in the top panels, the tracer particles are advected inward and spread more slowly than in the MAD case. The bottom panels similarly reflect this slower evolution, as the inferred advection velocity is closer to zero.
}
\label{fig:SANE_0p9_advection}
\end{centering}
\end{figure*}

The evolution of the spatial distribution function of the tracer particles contains signatures of both ordered inflow and radial mixing. We can obtain effective measures of inflow speeds and mixing rates by quantifying both the inward drift of the mean radius and the growth of the distribution width. Idealized models of mass transport often invoke the advection-diffusion equation, which for a quantity $C(r; t)$ can be written as
\begin{align}
\partial_t C = \partial_r \left( K \, \partial_r C - v \, C \right) + s
\end{align}
where $v$ is the net inward advection speed, $K$ is an effective radial diffusivity, and $s$ accounts for sources or sinks of particles (e.g., pair creation or absorption at the event horizon). For a purely diffusive, stationary process, $K$ would remain constant in time and the variance of the distribution would grow linearly with time $\sigma^2 \sim t$. In practice, however, there is no guarantee that the spreading is strictly diffusive. If particles followed ballistic trajectories, the spread would grow as $\sigma^2 \sim t^2$ and appear to obey a ``super-diffusive’’ scaling. Note, however, that $\sigma \sim t$ by itself is not sufficient to demonstrate ballistic motion: what matters is the underlying behavior of individual particles.

\subsection{Radial advection}
\label{sec:radial_advection}

We first examine the coherent inward motion of tracer particles near the disk midplane (here restricted to particles within $\left| z \right| \le 0.5 GM/c^2$), which corresponds to the advection term in the fluid description. To understand radial trends in the flow, we independently analyze shells of width $\Delta r = 0.5\ GM/c^2$. For each particle that passes through a given shell, we plot its radial trajectory as a function of time relative to the time that it passed through the shell. Figure~\ref{fig:MAD_0p9_advection} shows the accumulated trajectories of all particles passing through the radial shell $10 \leq r < 10.5\ GM/c^2$ in the high-resolution, high-spin prograde MAD simulation. The top panel shows how the radial distribution of particles passing through that radial shell evolves with time. It answers the question: Where are the particles before they pass through the shell and where do they end up afterwards? The solid white curve marks the mean trajectory of the accreting plasma, and the dashed curves denote the $\pm 1\sigma$ dispersion around the mean.

The mean trajectory in the top panel is smooth and monotonic, suggesting a clear net drift toward smaller radii. The spread of the distribution about the mean is nearly symmetric and Gaussian, suggesting that the radial motion can be decomposed into a combination of coherent inward flow and stochastic broadening. The bottom panel compares the tracer-inferred advection velocity, determined from the slope of the mean trajectory, to the radial fluid velocity measured directly from the GRMHD variables ($v^r = u^r/u^t$). The close agreement between the two confirms that the tracers faithfully track the bulk inflow of mass. Deviations from perfect alignment are small, and they may reflect differences between the net flow speed and a mass-weighted estimate of the infall velocity. We are unable to recover the dynamics of the tracer particles when they are very close to the event horizon since velocities cannot be inferred for particles that pass through the event horizon before the subsequent time step.

Figure~\ref{fig:SANE_0p9_advection} shows the same analysis for the equivalent SANE simulation. As in the MAD case, the SANE flow displays a smooth inward drift accompanied by a growing spread; however, the magnitude of $v^r$ in this radial bin is much smaller than its MAD counterpart, and the mean trajectory advances inward more slowly. Closer to the black hole, the SANE inflow also accelerates much more steeply than in the MAD case. Consistent with the MAD results, the tracer-derived advection speeds agree closely with the Eulerian fluid velocities, confirming that tracers reliably capture the bulk inflow dynamics.

\begin{figure*}
\begin{centering}
\includegraphics[width=\textwidth]{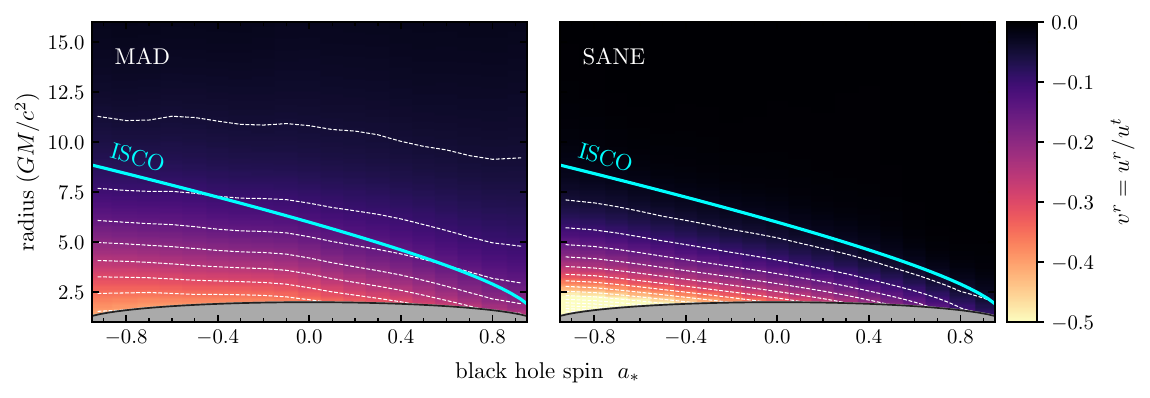}
\caption{
Radial velocity $v^r = u^r/u^t$ of the accreting plasma in the equatorial plane for GRMHD simulations with black hole spins $\bhspin = -0.9$ to $0.9$. Each panel shows time-averaged, axisymmetrized velocities over the interval $15,\!000-40,\!000\ GM/c^3$ from the $8/8$ resolution simulations. White dashed contours indicate $v^r = -0.05, -0.1, -0.15, \ldots$ in uniform spacing. The event horizon is denoted by the gray filled region, and the innermost stable circular orbit (ISCO) is plotted as the blue line. Both MAD and SANE flows exhibit trends with spin, with more rapid infall as spin decreases toward $\bhspin = -0.9$. MAD accretion flows exhibit stronger radial inflow at large radii, while SANE disks show a more rapid increase in $|v^r|$ with decreasing radius. Retrograde SANE flows in particular exhibit steep radial velocity gradients and enhanced infall near the black hole compared to MAD flows.
We use the $8/8$ resolution simulations to provide a more dense sample in spin and use fluid velocities rather than tracer-based measurements since the $8/8$ simulation output cadence is too long to robustly follow particle trajectories close to the event horizon. The velocities inferred from the tracers and the fluid data agree away from the horizon and are consistent across resolution.
}
\label{fig:inflow_velocities_grmhd}
\end{centering}
\end{figure*}

The unimodal shape of the displacement distributions in both MAD and SANE states indicates that a single bulk velocity captures the net radial motion at each radius. We see no evidence for strongly time-dependent behavior or multi-stage behavior, as might be expected if the flow alternated between phases of inflow and outflow (and especially in MAD disks, where one might anticipate large-scale outbursts). The smoothness of the mean trajectories and the symmetry of the spreads further suggest that turbulence and other mixing processes, though clearly active, do not overwhelm the coherent inward drift. In the MAD case, small but systematic deviations between tracer-based velocities and those inferred directly from the fluid highlight that the Eulerian fluid velocity profile need not exactly coincide with the mass-weighted average velocity traced by particles.

Since the tracer-derived velocities agree closely with the Eulerian fluid velocities, we look at the average fluid inflow velocity profiles across spins in Figure~\ref{fig:inflow_velocities_grmhd}. MAD flows begin significant inward acceleration at comparatively larger radii, with only modest variation across black hole spin. By contrast, SANE disks remain nearly circular until smaller radii, where they undergo sharper inward acceleration: the radius at which $v^r$ reaches $-0.05$ lies close to the innermost stable circular orbit (ISCO), especially for increasingly prograde spins. The spin dependence is also more pronounced in the SANE case, where inflow velocities in retrograde flows in SANE disks begin to exceed those of their MAD counterparts.

The absence of a sharp inflow transition near the ISCO in MAD flows highlights the different mechanisms that regulate the disk dynamics at horizon scales. In MADs, large-scale magnetic fields and the strong stresses they exert dominate angular momentum transport, dampening the thin-disk-like plunging signature and driving faster net inflow. As a result, MADs show larger $\| v^r \|$ and more rapid depletion of gas at small radii. By contrast, SANEs lack strong magnetic torques, so their inflow speeds increase more gradually and remain more closely tied to orbital dynamics, with a steeper rise in $v^r$ only inside of the ISCO. In both regimes, however, the classical pluging region is effectively smeared out by disk thickness and turbulence, and inflow velocities increase smoothly all the way to the horizon.

\subsection{Radial mixing}
\label{sec:radial_mixing}

We now turn to mixing in the disk, where turbulence, shocks, shear flows, and other instabilities cause initially co-located fluid elements to drift apart over time. Mixing randomizes the spatial ordering of plasma and redistributes angular momentum, energy, and magnetic flux independent of the mean inflow. Unlike coherent advection, mixing is inherently stochastic, and by reshuffling conserved quantities it can significantly alter the transport of conserved quantities even when the average inward velocity remains unchanged.

\begin{figure*}
\begin{centering}
\includegraphics[width=\textwidth]{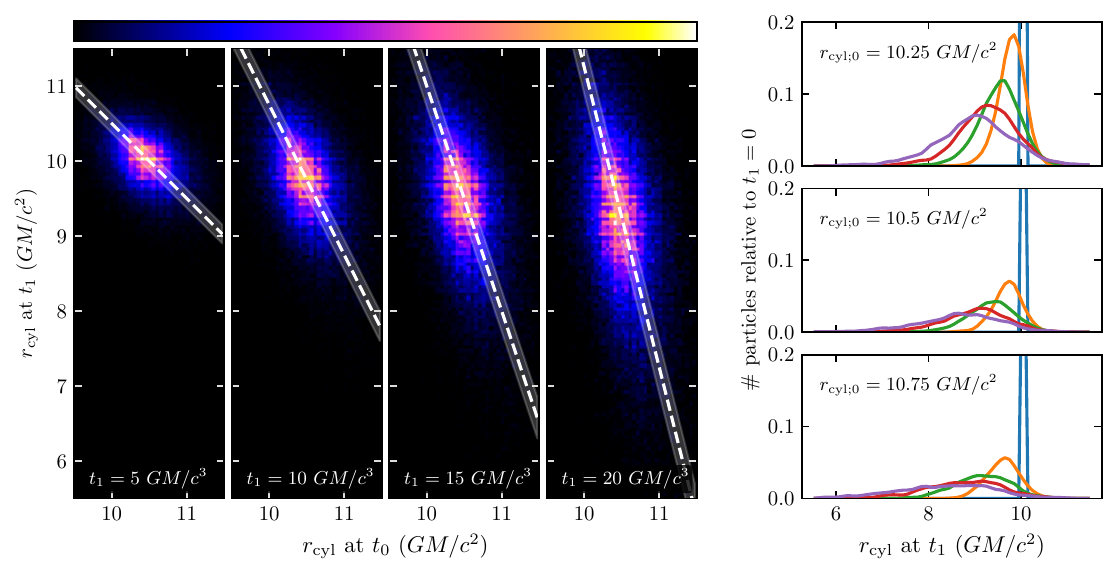}
\caption{
Evolution of radial position for particles that pass through the radial bin $10.2 \leq r_{\rm cyl} < 10.3\ GM/c^2$, aligned such that $t = 0$ corresponds to that crossing. Data computed for the high-spin $\bhspin = 0.9$ MAD simulation shown in Figure~\ref{fig:MAD_0p9_advection}.
Left: Distribution of particle positions in cylindrical radius $r_{\rm cyl}$ at time $t_0 = -5\ GM/c^3$ (horizontal axis) compared to $t_1 = 5, 10, 15, 20\ GM/c^3$ (vertical axis). The white-shaded band marks the kinematically allowed region for particles moving on ballistic trajectories between $t_0$ and $t_1$ while passing through the reference bin. The widening of the distribution beyond this band indicates the presence of mixing and turbulent dispersion beyond simple advection.
Right panels: Evolution of the distribution function over time relative to the distribution at $t_1=0$ (blue; other colors at $t_1=5,10,15,20\ GM/c^3$) for three representative initial locations: $r_{\rm cyl} = 10.25$, $10.5$, and $10.75\ GM/c^2$ (corresponding to different vertical slices in the left panels). Each panel shows both coherent inward advection and broadening of the distribution function, indicating mixing over time.
}
\label{fig:mad_spreading_example}
\end{centering}
\end{figure*}

\begin{figure*}
\begin{centering}
\includegraphics[width=\textwidth]{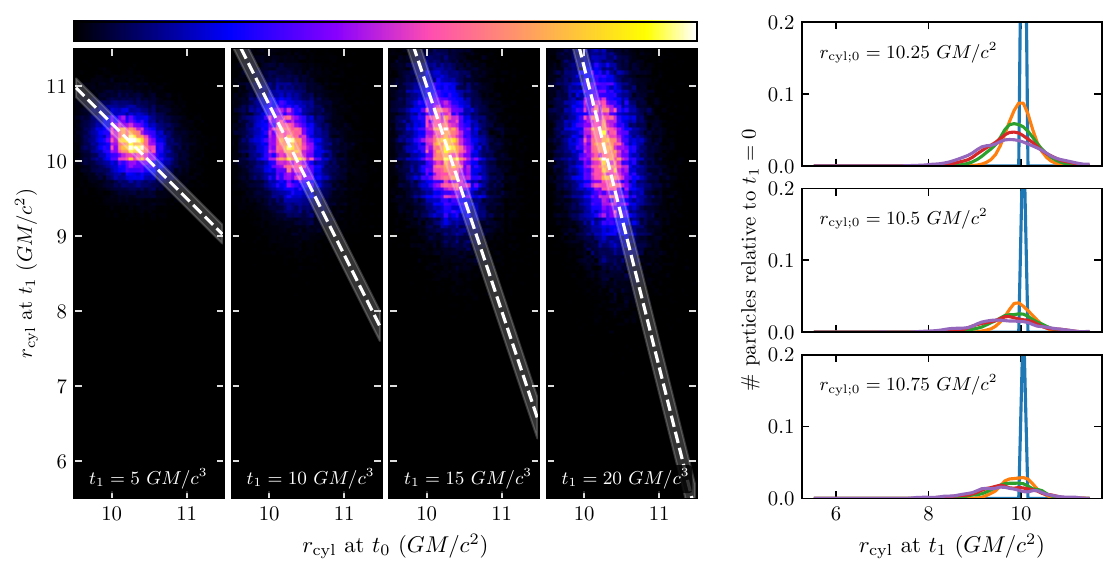}
\caption{
Same as Figure~\ref{fig:mad_spreading_example}, but for the high-spin $\bhspin = 0.9$ SANE simulation shown previously in Figure~\ref{fig:SANE_0p9_advection}. The tracer distribution shows evidence of both inward advection and spreading, although both rates are slower compared to the MAD flow.
}
\label{fig:sane_spreading_example}
\end{centering}
\end{figure*}

\begin{figure*}
\begin{centering}
\includegraphics[width=0.95\textwidth]{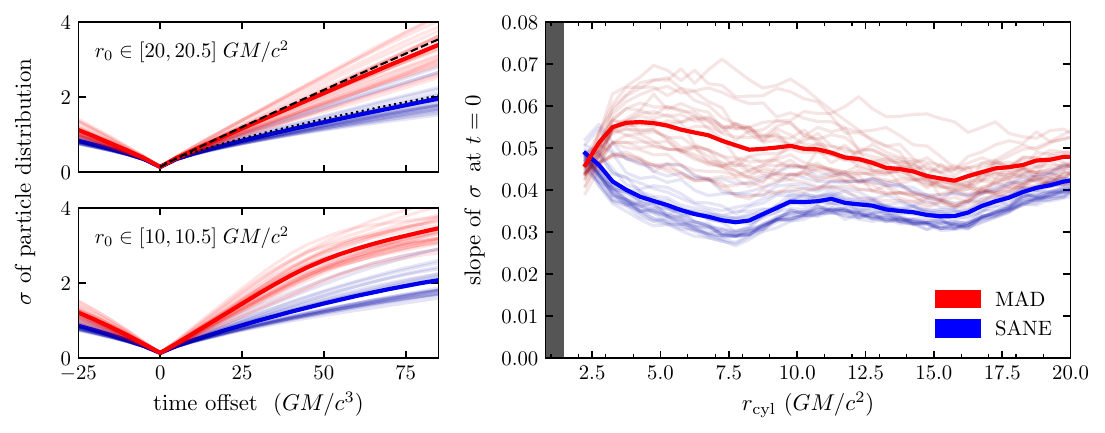}
\caption{
Temporal and spatial distribution of radial mixing in the same high-spin MAD (red) and SANE (blue) simulations shown in earlier figures. 
Left: Growth of tracer particle distribution width $\sigma$ over time, starting from a population of particles that occupies a particular radial shell ($r_0$; top/bottom) at time offset $=0$. The spread of the distribution grows super-diffusively at early times for both simulations, with average behavior over the later part of the simulation (bold lines) bracketed by analytic scalings $\sigma \propto t^{\,0.95}$ (dashed) and $\sigma \propto t^{\,0.75}$ (dotted). Fainter lines show results from individual time segments. At later times, changes in the width reflects mixing history from smaller radii and does not approach a true steady state. Right: Initial growth rates of $\sigma({\rm time\ offset})$, measured from the local slope of $\sigma$ shortly near the point where time offset $=0$, shown as a function of radius. Measurements near the horizon are excluded where truncation of the distribution prevents reliable recovery (see Appendix~\ref{sec:appendix_fitting}).
}
\label{fig:spreading_traces_highspin}
\end{centering}
\end{figure*}

We measure mixing by isolating the spreading component of the tracers distribution function, which is not captured by the mean drift. Specifically, we measure the time-dependent standard deviation $\sigma$ of the radial distribution function of particles that share a common reference point in their trajectories (see Figures \ref{fig:MAD_0p9_advection} and \ref{fig:SANE_0p9_advection}). In idealized turbulent diffusion, $\sigma\sim t^{1/2}$, but realistic accretion flows are shaped by a combination of turbulence, coherent shear, instabilities, and transient large-scale structures. As a result, we anticipate departures from simple diffusive scaling, with the growth rate of $\sigma$ depending both on radius and on the underlying accretion state.

The bulk inflow of the plasma gives rise to two important caveats. First, because the tracer cohort drifts inward while it spreads, the growth of $\sigma$ reflects not only the local mixing properties but also the changing flow properties with decreasing radius. Second, as particles cross the event horizon and are removed from the sample, the measured distribution becomes truncated, which limits our ability to probe the asymptotic long-time limit. The early-time scaling of $\sigma(t)$ thus provides the clearest window into the intrinsic local mixing process.

However, $\sigma(t)$ is not a pure diagnostic of turbulent diffusion. Its evolution is shaped by both genuinely stochastic spreading as well as a purely kinematic effect: tracers that cross the same reference bin with different radial velocities will diverge in position even if each follows a perfectly ballistic path. This broadening due to velocity-shear is not ``mixing’’ in the strict diffusive sense, but it still redistributes material radially and therefore contributes to effective mixing from a mass-transport perspective.

To assess whether genuine diffusion-like mixing occurs, we track how particles that pass through a given radial shell are redistributed over short time intervals. The left panels of Figure~\ref{fig:mad_spreading_example} show the initial positions of tracers at $t_0 = -5\ GM/c^3$ versus their positions at later times $t_1 = 5, 10, 15, 20\ GM/c^3$. The initial position serves as a proxy for the radial velocity of the particle when it crosses the target shell. In the absence of diffusive motion, the particles would remain confined within the white-shaded diagonal bands that mark the kinematically allowed displacement range between the two times.

Any systematic broadening beyond the ballistic band is evidence for transport processes that cannot be explained by coherent motion alone. The right-hand panels of Figure~\ref{fig:mad_spreading_example} show the time evolution of tracer subpopulations selected by their initial positions. Each group drifts inward coherently as its distribution widens, yielding a progressively broader spread over time. This combination of ordered advection with stochastic broadening is the hallmark of mixing that is at least partially diffusive in character. The corresponding SANE results, shown in Figure~\ref{fig:sane_spreading_example}, reveal qualitatively similar trends: tracers consistently spread beyond their ballistic limits, but the net advection is slower and the overall rate of broadening is reduced compared to the MAD case.

We quantify the magnitude of radial spreading by tracking the standard deviation of the radial distribution function over time. This measure captures not only diffusive broadening but also any additional divergence arising from other processes and therefore provides a single diagnostic of how material disperses over time. Figure~\ref{fig:spreading_traces_highspin} shows the time evolution of $\sigma(t)$ for the same high-spin MAD and SANE simulations. Each faint curve corresponds to a cohort of tracers that cross the fiducial radial bin within a $1000\ GM/c^3$ window. The bold curves show the average trend, computed over the same long time interval used in the radial-velocity analysis ($t = 15,\!000-40,\!000\ GM/c^3$).

\begin{figure}
\begin{centering}
\includegraphics[width=\linewidth]{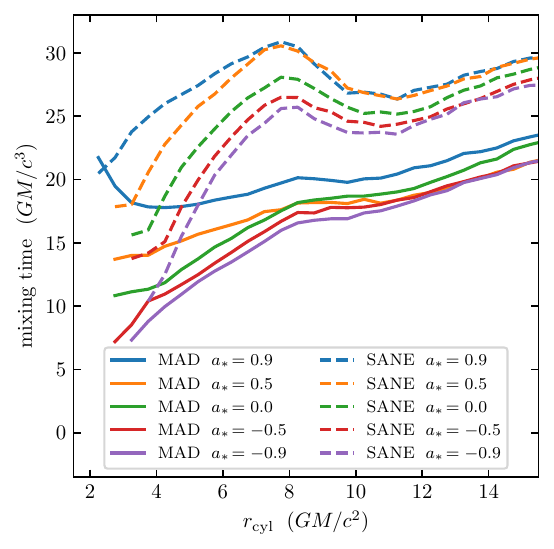}
\caption{
Mixing times as a function of radius for matter within the midplane in the high resolution simulations. We define mixing time as the time required for a narrow ring of tracer particles to spread radially to a standard deviation of $1\ GM/c^2$, assuming a constant rate of spread equal to the initial rate near time offset $\Delta{}t = 0$. Mixing is consistently faster in MAD simulations than in SANE simulations. Retrograde flows also tend to exhibit shorter mixing times than their prograde counterparts. Mixing times at small radii generally decrease as matter approaches the black hole, although this trend is not strictly monotonic across all models and may reflect the plasma state being ``frozen'' at larger radii before plunging toward the black hole.
}
\label{fig:mixing_times}
\end{centering}
\end{figure}

In both simulations, the early-time growth of $\sigma$ is superdiffusive, increasing more rapidly than the $t^{1/2}$ scaling expected for pure diffusion. Power-law fits yield $\sigma(t) \propto t^{\,0.95}$ for the MAD flow and $\sigma(t) \propto t^{\, 0.75}$ for the SANE one. At all times, the MAD curves lie above the SANE ones, suggesting that radial dispersion is amplified when the magnetic fields are stronger. This enhanced mixing in MAD flows is plausibly driven by magnetic interchange and other instabilities associated with magnetically dominated regions.

We compare and contrast mixing across our simulation library in terms of a characteristic mixing time, which may vary with radius. We define the mixing time as the interval required for the standard deviation in the tracer distribution function to increase by $\Delta \sigma = 1\ GM/c^2$, assuming a constant growth rate equal to the initial slope of $\sigma(t)$. Figure~\ref{fig:mixing_times} shows the resulting curves for the high-resolution simulations. (As before, the results agree across resolution outside of the innermost radii, but temporal finite sampling and the event-horizon truncation limit complicate the recovery of robust statistics closer in.) In both MAD and SANE states, mixing times typically decrease toward smaller radii, consistent with shorter dynamical timescales and stronger mixing near the black hole. Notice, however, that the mixing time in the high-spin prograde MAD flow increases very close to the black hole. This behavior may reflect the spatial structure of the plasma being ``frozen into'' strands of accreting material as they enter the plunging region close to the event horizon.  At all radii, however, MAD flows mix more efficiently than SANEs, reflecting the role of magnetic interchange and enhanced stresses in magnetically arrested disks. Retrograde configurations also tend to exhibit shorter mixing times than their prograde counterparts, paralleling the faster advection speeds measured above.

\section{Disk Evolution}
\label{sec:disk_evolution}

The use of Lagrangian tracer particles allows us to track the trajectory of matter from its initial position to its eventual accretion through the event horizon and to study the combined influences of coherent inflow, turbulent mixing, and long-term disk depletion. In a purely advective, laminar flow, fluid elements would follow infalling streamlines at the local advection speed, and matter accreting at some time would be sourced primarily from a single radius. In contrast, the turbulent and diffusive disks studied here source matter from a wider radial distribution, as fluid parcels from separated radii mix and wander stochastically before reaching the black hole. We use tracer particle histories to directly link the position of each accreting element at the horizon back to its location in the initial torus.

This mapping provides a practical estimate of the inflow time, defined as the time required for material from a given initial location to reach the black hole. It also enables a direct comparison of how inflow and draining proceed across different accretion states, spins, and initial conditions, offering insight into the structure and evolution of each accretion flow. We begin by comparing the time-dependent evolution of the accretion source in the long-duration, high-spin MAD and SANE simulations, then explore systematic differences in accretion behavior across models, before connecting these trends to the rate at which the initial mass reservoir in the tori is depleted over time.

\subsection{Inflow and the source of the accreting plasma}

\begin{figure*}
\begin{centering}
\includegraphics[width=\textwidth]{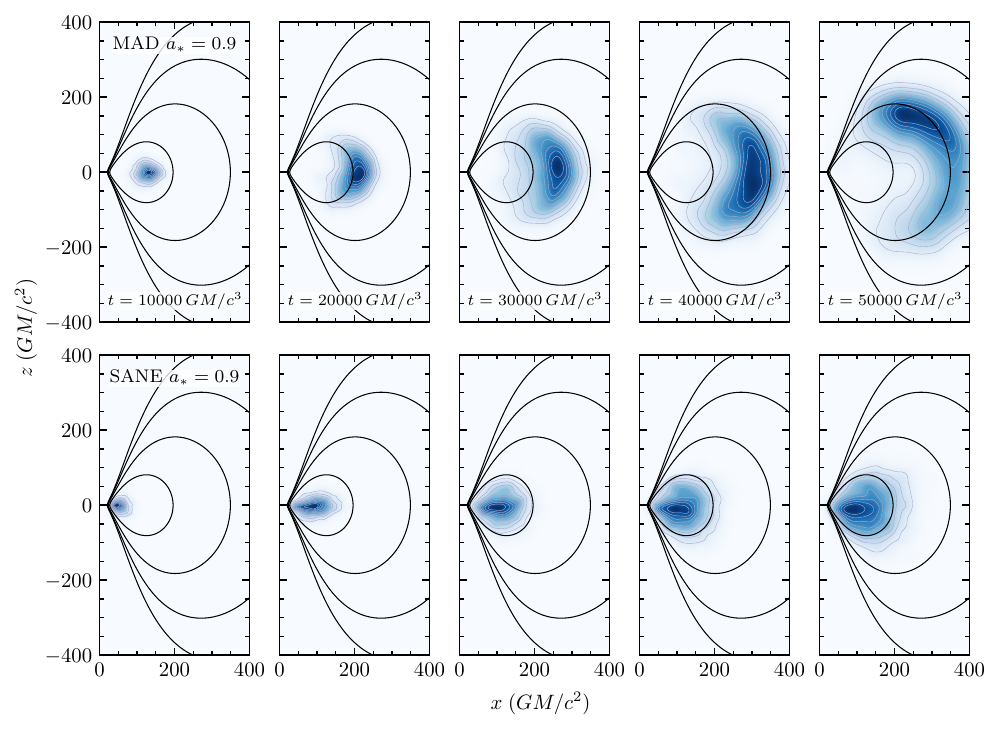}
\caption{
Initial positions of matter accreting through the event horizon for high-spin $\bhspin = 0.9$ MAD (top row) and SANE (bottom row) simulations. Each panel shows an azimuthal slice in Kerr–Schild coordinates, colored by a kernel density estimate (KDE) of all tracer particles that accrete within $1,\!000\ GM/c^3$ of the listed time. Light gray contours mark KDE levels at $0.1, 0.2, \ldots, 0.9$ of the maximum value. Black contours show the initial mass distribution, drawn at even logarithmic intervals from $1\%$ to $100\%$ of the maximum initial density. All values are integrated $d \phi$. Early accretion is dominated by particles within a narrow midplane wedge; over time, a larger fraction of the disk contributes. The MAD simulation shows broader, more radially extended source regions than the SANE simulation (the asymmetric behavior in the $t=50,\!000\ GM/c^3$ panel is a random transient, see the next figure). The matter source in the SANE also begins to spread vertically around $t \approx 20,\!000\ GM/c^3$, coinciding with accretion from the initial mass maximum. Finally, notice that in the SANE flow material is drawn from throughout the draining region, while in the MAD flow the inner disk drains quickly and the dominant source region moves outward.
}
\label{fig:accretion_source_rphi}
\end{centering}
\end{figure*}

Figure~\ref{fig:accretion_source_rphi} shows how the typical origin of accreted matter in the long-duration MAD and SANE flows evolves with time, reflecting the outward propagation of the inflow region and providing a measure of how long it takes material from a given radius to accrete. Early accretion in both MAD and SANE flows is dominated by matter drawn from a narrow wedge near the midplane. In the MAD flow, however, the dominant source radius increases rapidly as the inner disk is depleted, resulting in a broader, more radially extended origin distribution that marches steadily outward. The asymmetry in the MAD at $t=50,\!000\ GM/c^3$ is a short-lived random transient (see the same time in the next figure). In contrast, the SANE flow retains persistent contributions from smaller radii even as accretion samples an increasing fraction of the torus. 

\begin{figure*}
\begin{centering}
\includegraphics[width=\textwidth]{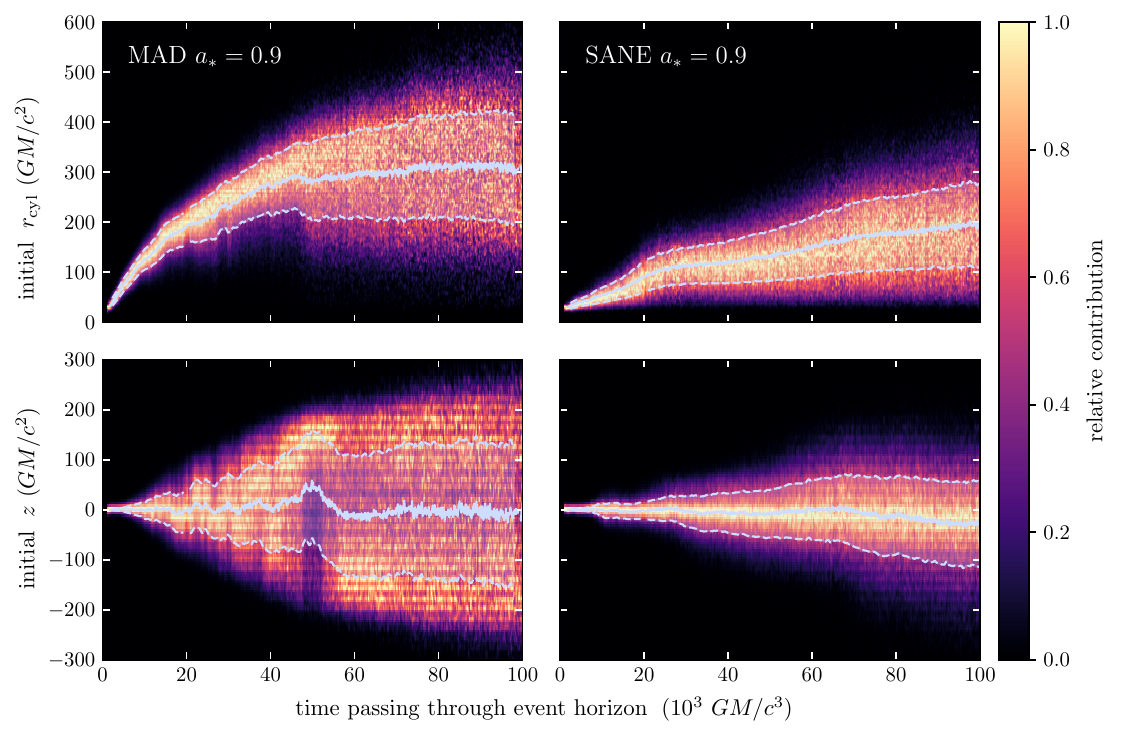}
\caption{
How the initial position of the accreting matter changes over time in the high-spin $\bhspin = 0.9$ MAD (left) and SANE (right) simulations.
Top: Initial cylindrical radius $r_{\rm cyl}$ of accreting matter (vertical axis) versus the time it crosses the event horizon (horizontal axis), normalized within each time bin. Each horizontal slice indicates the time when matter at a given $r_{\rm cyl}$ in the initial torus accretes through the event horizon. In the MAD flow, small radii are depleted more rapidly, and the characteristic source radius grows steadily with time; by $t \approx 50,\!000\ GM/c^3$, the source region lies at large radii and shows little secular evolution. In the SANE simulation, once accretion begins from the radius of maximum initial mass, the growth rate of the mean source radius slows. Solid and dashed lines indicate the mean and standard deviation of each distribution, respectively.
Bottom: Same as above but showing the initial height $z$ above the midplane. The width of the $z$ distribution reflects both turbulent vertical mixing and the radial variation of the initial torus scale height (see Figure~\ref{fig:grmhd_overview}).
}
\label{fig:ratios2d}
\end{centering}
\end{figure*}

Figure~\ref{fig:ratios2d} shows the distribution of the initial positions of accreted particles as functions of accretion time for the same two long-duration simulations considered in Figure~\ref{fig:accretion_source_rphi} along with fits to the distribution mean and standard deviation for each accretion time.
In the MAD flow, the mean source radius grows monotonically as inner regions are drained. By $t \approx 50,\!000\ GM/c^3$ the accreting material originates almost entirely from large radii, and the distribution of the source positions remains relatively constant. By contrast, in SANEs the outward drift of the mean is slower, and the distribution broadens more than it shifts, reflecting sustained contributions from a wide range of radii. Vertically, both flows exhibit increasing spread with time, consistent with turbulent mixing and the radial variation in torus scale height, but the effect is more pronounced in MAD, where rapid radial depletion exposes vertical structure more quickly. 

These trends mirror the advection and mixing behavior described in Section~\ref{sec:advection_mixing}. In MADs, shorter mixing times and stronger mass transport mean that the inner radii are depleted more quickly and the source region quickly moves outward. By contrast, in SANEs the mean source radius grows more slowly and the origin distribution broadens rather than shifts. Because the draining rate is also slower, SANEs effectively have more time to mix before their inner reservoirs are exhausted, allowing them to approach a quasi-equilibrium state more readily than MADs, where rapid inflow continually pushes the accretion source outward.
    
\subsection{Comparison across models}

\begin{figure*}
\begin{centering}
\includegraphics[width=0.9\textwidth]{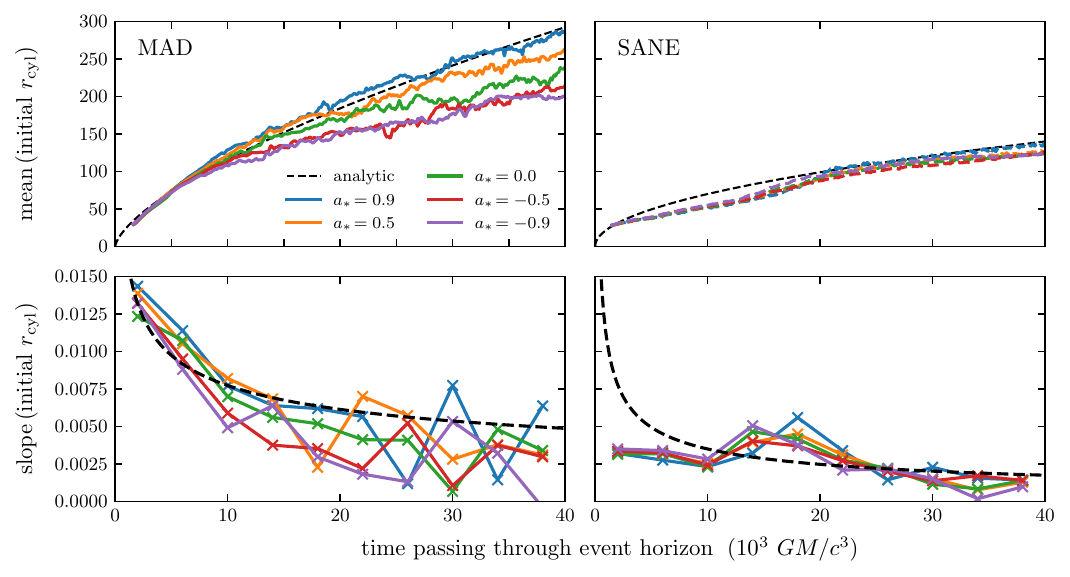}
\caption{
Evolution of the initial source location for matter accreting through the event horizon across simulations.
Top: Mean initial cylindrical radius $r_{\rm cyl}$ of accreting material as a function of time for MAD (left) and SANE (right) models. In MADs, the mean source radius moves outward rapidly as the inner disk drains quickly; in SANEs, the shift is slower and more gradual.
Bottom: Instantaneous rate of change of the mean source radius, computed from the slopes of the curves in the top panels. MADs begin with steep slopes, reflecting rapid early draining, while SANEs evolve more slowly. All SANE models show a feature near $t \approx 20,\!000\ GM/c^3$, corresponding to when accretion begins to tap the initial peak of the torus mass distribution (this behavior is seen across resolutions). Reference power-law scalings are overplotted: $r_{\rm cyl} = 0.25\,t^{2/3}$ for MADs and $r_{\rm cyl} = 0.7\,t^{1/2}$ for SANEs.
The continued outward drift of the mean source radius reflects progressive depletion of the initial torus rather than a stationary inflow. Nonetheless, the well-defined draining trends in both MAD and SANE models enable a rough estimate of an effective inflow time as a function of radius, with MADs exhibiting systematically shorter timescales than SANEs at comparable radii.
}
\label{fig:source_distribution_evolution_subsample}
\end{centering}
\end{figure*}

We apply the same origin-mapping procedure to each simulation in our survey to compare how the source region evolves for different magnetic configurations and black hole spins. We also compare across different resolutions to confirm the robustness of our results. 
Figure~\ref{fig:source_distribution_evolution_subsample} summarizes how the mean initial cylindrical radius of accreting material evolves with time as a function of magnetization (MAD vs.~SANE) and black hole spin. Only the high-resolution models are plotted for clarity, but the standard-resolution models exhibit the same trends in behavior. In MAD flows, the source radius increases rapidly as the inner disk is depleted. This early-time rise is nearly universal across MAD cases until the source reaches $r_{\rm cyl} \approx 100\ GM/c^2$, after which the curves diverge, with the source region increasing more rapidly for larger values of $\bhspin$.
These spin-dependent differences partly reflect intrinsic draining rates but also arise from variations in the initial torus size (see Figure~\ref{fig:grmhd_overview}). A simple power-law fit to the early-to-intermediate MAD phase suggests $r_{\rm cyl} \propto t^{2/3}$, though the derivative curves in the lower panels confirm that this scaling is only approximate.

SANE accretion flows exhibit qualitatively different behavior. For $t \lesssim 20,\!000\ GM/c^3$, the mean source radius grows slowly until it reaches approximately $r_{\rm cyl} \approx 50-60\ GM/c^2$, after which point the mean source radius grows rapidly to approximately $r_{\rm cyl} \approx 100\ GM/c^2$. At $t\approx 20,\!000\ GM/c^3$, the curves exhibit a distinct break and the slope flattens. A plausible explanation for this behavior is that once this inner source is tapped, the accretion stream remains supplied from a broad range of radii, but the relative contribution from larger radii is limited since there is less mass there, and the smaller‐radius material dominates the mass budget. At late times, the growth can be approximated by $r_{\rm cyl} \propto t^{1/2}$, although this scaling fails to capture the pre-break regime.
The bottom panels of Figure~\ref{fig:source_distribution_evolution_subsample} show the contrast in draining rates and the quality of the fit: MAD models start with steep $d r_{\rm cyl}/d t$ slopes that decay rapidly, while SANE slopes remain lower and more uniform aside from the break near $t\approx 20,\!000\ GM/c^3$.

\begin{figure}
\begin{centering}
\includegraphics[width=\linewidth]{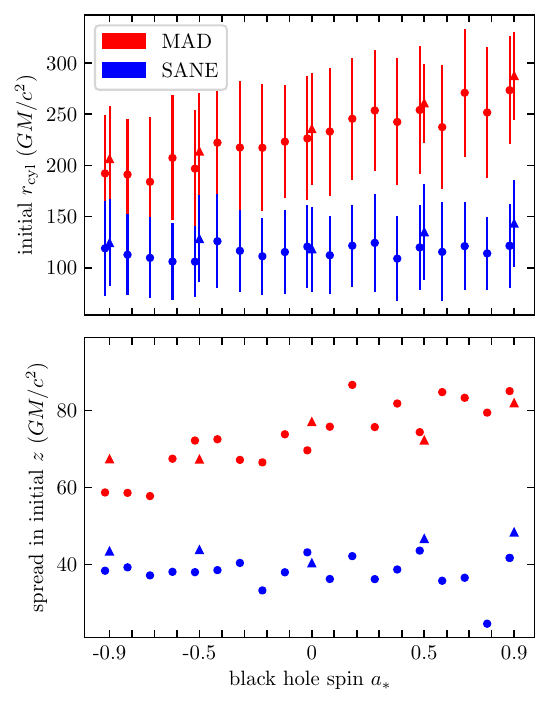}
\caption{
Initial location of material accreting through the event horizon near $t = 40,000\ GM/c^3$ across all simulations. MAD simulations are shown in red and SANE simulations in blue, with marker shape indicating resolution (circles for $8/8$ and triangles for $16/8$, offset for clarity).
Top: Mean and standard deviation of the initial $r_{\rm cyl}$ distribution of accreting material. MAD simulations show systematically larger mean radii and broader spreads than their SANE counterparts. The mean radius for MAD models also increases from strongly retrograde to strongly prograde spin, although this may partly reflect differences in the initial torus size combined with larger accretion rates (see Figure~\ref{fig:grmhd_overview}).
Bottom: Standard deviation of the initial $z$-position for the same tracer populations. The trends in this vertical spread correlate with the initial $r_{\rm cyl}$ shown in the top panel: As the initial radial position increases, the accretion flow is able to draw from a larger range of vertical positions, since the initial torus increases in height with radius.
} 
\label{fig:origin_fit_vs_model}
\end{centering}
\end{figure}

Figure \ref{fig:origin_fit_vs_model} summarizes the differences in the source evolution across all simulations in our library at a common reference time of $t = 40,\!000\ GM/c^3$, by which point all runs have evolved beyond the initial transient but have not yet fully drained the torus. The top panel shows the mean and standard deviation of the initial cylindrical radius $r_{\rm cyl}$ of accreting material. MAD simulations (red) consistently exhibit larger mean radii and broader spreads than their SANE counterparts (blue). These differences persist across spin values and resolutions, reinforcing the conclusion that MAD accretion proceeds more rapidly, likely due to enhanced angular momentum transport and stronger mixing.

The influence of black hole spin is more nuanced. In the SANE flows, spin appears to have negligible impact on the source radius distribution. In MAD flows, however, the mean source radius tends to increase with spin, growing from highly retrograde ($\bhspin = -0.9$) to highly prograde ($\bhspin = 0.9$). As shown in Figure~\ref{fig:source_distribution_evolution_subsample}, the differences across spin become apparent primarily after the mean source radius reaches approximately $100\ GM/c^2$, which is where the initial mass distribution function in the initial tori begin to diverge significantly. We return to this point in the next subsection, but note that the variation in accretion rate between models suggests that this effect is not solely attributable to differences in initial conditions. Nevertheless, this difference in draining behavior is a practical consideration when performing GRMHD simulations using similar initial tori.

\subsection{Disk draining and the mass budget}

\begin{figure}
\begin{centering}
\includegraphics[width=\linewidth]{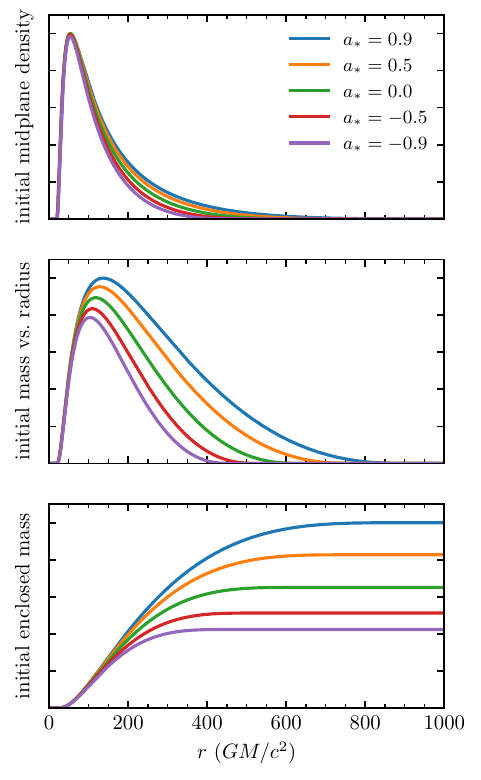}
\caption{
Initial distribution of coordinate-frame torus mass ($\rho u^t \sqrt{-g}$) as a function of black hole spin. Top: Midplane mass density versus radius. All tori peak within a narrow range, $r \approx 53-56\ GM/c^2$, but those in higher-spin spacetimes extend farther outward. Middle: Radial distribution of total mass, integrated over $\theta$. Because the more extended tori place more material at large radii, the peak radius of the mass distribution ranges from $r \approx 105-135\ GM/c^2$ across spins. Bottom: Cumulative enclosed mass versus radius. The $\bhspin=0.9$ torus contains nearly $2.5\times$ as much mass as the $\bhspin=-0.9$ case. The half-mass radius also varies widely, from $r \approx 145\ GM/c^2$ to $r \approx 235\ GM/c^2$.
} 
\label{fig:initial_mass_distribution}
\end{centering}
\end{figure}

\begin{figure}
\begin{centering}
\includegraphics[width=\linewidth]{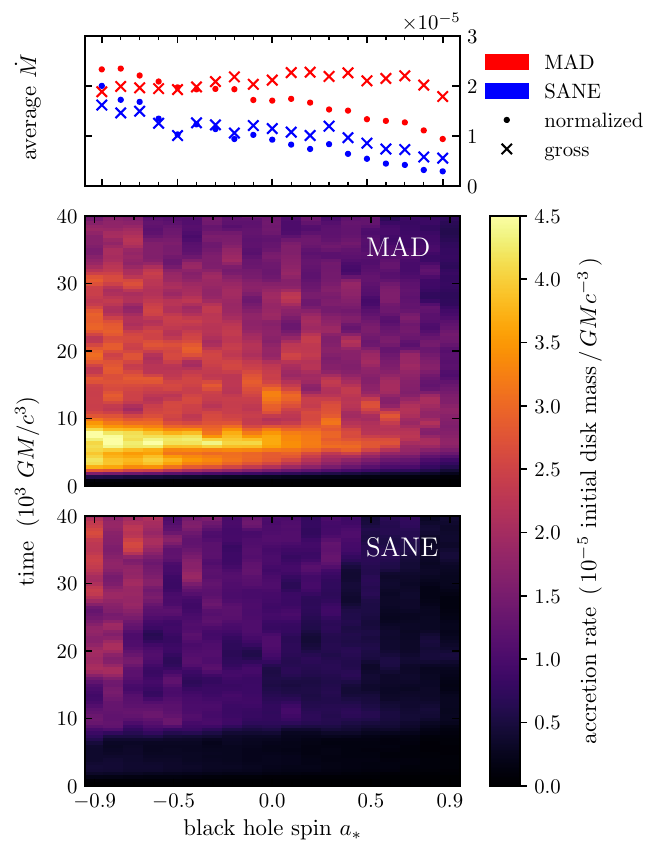}
\caption{
Accretion rate as a function of time across simulations.
Center and bottom: Heatmaps of accretion rate (color) versus time for each standard-resolution simulation. Rates are normalized by the total initial torus mass (see Figure~\ref{fig:initial_mass_distribution}), so color corresponds to the fraction of disk mass crossing the horizon per $GM/c^3$. MAD simulations accrete significantly faster than their SANE counterparts and show a pronounced transient burst between $5,\!000-10,\!000\ GM/c^3$, especially in retrograde, high-spin cases.
Top: Time-averaged accretion rates measured over $15,\!000-40,\!000\ GM/c^3$. Filled circles show normalized rates, while x-marks show the corresponding gross (unnormalized) values. Although the gross accretion rate in MADs peaks near low prograde spins, the normalized rate increases steadily toward $\bhspin=-0.9$ because the initial torus mass decreases in that direction.
} 
\label{fig:accretion_rates_detail}
\end{centering}
\end{figure}

The outward motion of the accretion source region described above is closely tied to the progressive depletion of the finite-mass torus. Although mass supply in the real world will be continuously sourced from large radius, in our finite-mass tori, mass is lost as the inner disk accretes and the locus of accretion necessarily shifts outward (other initial and boundary conditions like those considered by \citealt{cho_2025_largelong} or \citealt{olivares_2023_transonic} and different methods to model the incoming gas like in \citealt{guo_2023_towardhorizonscale,guo_2025_cycliczoom,cho_2025_multizone} and \citealt{ressler_2018_stellarwinds,ressler_2020_madstellarwinds,hopkins_2024_fire} may be less subject to disk draining). 
Figure~\ref{fig:initial_mass_distribution} shows that the initial tori differ substantially in radial extent and total mass as a function of spin. For example, the $\bhspin=0.9$ torus contains nearly $2.5\times$ more mass than the $\bhspin=-0.9$ case, with half the mass located as far out as $r \approx 235\ GM/c^2$ in contrast to $r \approx 145\ GM/c^2$ for $\bhspin = -0.9$.
While the draining behavior in our simulations is influenced by the choice of initial condition, it offers a quantitative baseline that directly reflects the efficiency of angular momentum transport and mixing in the accretion flow: for a given magnetic state, spin, and resolution, we can estimate the time required to remove a specified fraction of mass at a given radius.

The rate at which accretion through the event horizon drains the torus is shown in Figure~\ref{fig:accretion_rates_detail}. After normalization by the initial torus mass, MAD simulations consistently drain more efficiently than their SANE counterparts. While the \emph{gross} accretion rates in MAD flows peak near intermediate prograde spins, the normalized rates increase steadily toward retrograde spins, reflecting the differences in the initial conditions. Together with the initial torus profiles shown in Figure~\ref{fig:initial_mass_distribution}, these rates help explain the late-time behavior of the accretion source centroids in Figures~\ref{fig:source_distribution_evolution_subsample} and~\ref{fig:origin_fit_vs_model}: the MAD flows accrete at similar rates early on, but once the accretion front passes the radii where the mass distribution peaks, their evolution begins to diverge since there is less material at large radii to feed the accretion. 

These accretion rates also allow us to estimate the total mass in the original torus. If we rescale the code density in the simulations using the physical accretion rates inferred for M87 in \citep{eht_m87_8}, the initial MAD and SANE tori for $\bhspin=0.9$ would contain $1 - 10$ and $10 - 100\ M_\odot$, respectively. This means that the total mass in the torus would be between $\approx 10^{-10} - 10^{-8}$ the mass of the central black hole, supporting the common argument that the influence of the accreting matter on the spacetime metric can be neglected.

\begin{figure*}
\begin{centering}
\includegraphics[width=\textwidth]{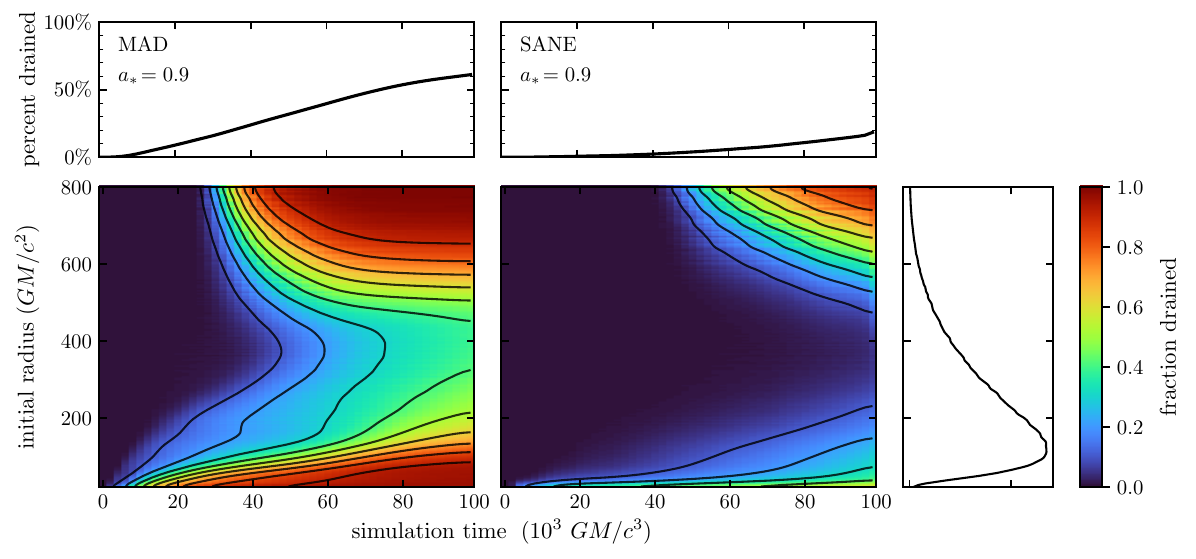}
\caption{
Disk draining in the long-duration, high-spin $\bhspin = 0.9$ MAD and SANE $8/8$ simulations.
Top: Fraction of the initial torus mass that has left the domain (either through the event horizon or via outflows) as a function of time.
Bottom left and bottom center: Fraction of the initial mass at each radius (vertical axis) that has exited the domain over time. Contours indicate the times at which $10\%, 20\%, \dots, 90\%$ of the initial mass at a given radius has been removed, highlighting how depletion rates vary across the disk.
Bottom right: Initial radial distribution of mass in the torus (same for MAD and SANE). The draining rate depends on both disk structure and magnetic configuration, with MAD systems depleting more rapidly at larger radii.
}
\label{fig:disk_draining}
\end{centering}
\end{figure*}

A more resolved view of the draining behavior is provided in Figure~\ref{fig:disk_draining}, which quantifies the position-dependent draining in the two long-duration simulations and provides a complementary means to identify the characteristic time for mass in the initial condition to be depleted. The timescales and trends in depletion across radius differ markedly between the two states. The top panels show the fraction of the initial torus mass that has left the computational domain over time, whether via accretion through the event horizon or outflow through the outer radial boundary. By $t \approx 74,\!000\ GM/c^3$, the MAD flow has lost half of its initial mass, while the SANE torus has drained only about $9\%$ over the same interval.

The bottom panels differentiate the draining behavior by the initial cylindrical radius. In the MAD, depletion proceeds rapidly at both small and large radii in the torus, with approximately half of the initial mass within $100\ GM/c^2$ being drained by $t = 50,\!000\ GM/c^3$. Matter initialized at large radii also leaves the domain, although since it leaves through the simulation boundary its depletion time is less well defined. Draining in the SANE flow proceeds much more slowly, with less than $30\%$ of the initial mass within $r_{\rm cyl} = 100\ GM/c^2$ drained even by $t=100,\!000\ GM/c^3$.

\section{Discussion}
\label{sec:discussion}

We have used Lagrangian tracer particles in \athenak to study mass transport in radiatively inefficient, advection-dominated black hole accretion flows. We produced a library of 48 GRMHD simulations spanning a range of black hole spins in both the magnetically arrested disk (MAD) and standard and normal evolution (SANE) magnetic flux states. We have used the tracer histories to characterize and quantify advection and stochastic mixing as well as to study inflow times and disk-draining behavior in simulations initialized with finite-mass tori.

In studying mass transport and mixing within the disk, we find:
\begin{itemize}[leftmargin=2em]
\item The distribution of tracer particles passing through a given radial shell is well described by a combination of net inflow and Gaussian-like broadening, akin to an advection-diffusion process.
\item MAD simulations exhibit systematically faster inflow than SANEs, and radial velocities increase toward the event horizon in all cases. The ISCO plays little role in MADs but leaves a clearer imprint in SANEs.
\item Decreasing the black hole spin (from high-spin prograde to zero-spin to high-spin retrograde) tends to increase the inflow velocity. High-spin SANE retrograde flows in particular exhibit the most rapid infall, with radial velocities rising steadily up to the event horizon.
This trend likely reflects the balance between magnetic pressure support and gravity and the diminished role of the ISCO in MAD flows, where magnetic pressure plays a larger role.
\item Turbulence leads to a strong stochastic spread of fluid elements, driving mixing within the disk. MADs mix more efficiently than SANEs, but all cases are superdiffusive, with tracer spread scaling roughly as $\sigma \sim t^{\,0.95}$ in MADs and $\sigma \sim t^{\,0.75}$ in SANEs for prograde $\bhspin=0.9$.
\item Mixing times are systematically shorter in MADs than in SANEs and decrease toward the event horizon, and prograde disks mix more slowly than retrograde ones. In the prograde MAD case, the mixing time \emph{increases} very close to the event horizon, which maybe explained by the plasma state being ``frozen'' at larger radii before plunging inward. 
\end{itemize}
We have also used tracer particles to reconstruct the origins of accreted matter and to track how turbulent disks evolve and drain over time. We find that:
\begin{itemize}[leftmargin=2em]
\item Turbulent mixing leads to accretion drawing material from a broad range of initial radii, in contrast to laminar flows where the accreting plasma would be sourced from a narrow radial band that moves outward with time. The effect is especially pronounced in MADs, where rapid depletion of the inner torus quickly drives the mean source radius outward.
\item In MADs, the mean source radius grows approximately as $r_{\rm cyl} \propto t^{2/3}$. In SANEs, the late-time behavior is closer to $r_{\rm cyl} \propto t^{1/2}$, with a distinct break near $t \sim 20,\!000\ GM/c^3$.
\item Spin effects are most apparent in MADs, where accretion from prograde disks comes from larger mean source radii at late times. However, these trends partly reflect differences in the initial mass distribution of the tori.
\item The finite mass of the initial torus strongly shapes accretion histories. Compact tori deplete more rapidly, especially in MADs, where signatures of draining are visible as early as $t \sim 10,\!000\ GM/c^3$.
\end{itemize}

Our work is subject to several important limitations. Our implementation means that the precision with which the tracer particles sample the spatial distribution of matter is limited to the grid scale of the simulation, and the details of the Monte Carlo particle evolution algorithm introduces an additional effective diffusivity. Although this treatment is appropriate for characterizing bulk transport, understanding accretion histories, and studying the statistics of mixing, it is not suitable for all physical applications and is limited by resolution. In addition, since particle histories are truncated at the event horizon, our ability to recover dynamics close to the event horizon is limited by our output cadence (see also Appendix~\ref{sec:appendix_fitting}). As with all GRMHD studies, finite resolution and numerical effects remain unavoidable sources of uncertainty.

There are a number of natural extensions to this work. Studying the trajectories of the particles that are ejected from the disk in outflows would help explain how mass loading and acceleration proceed in winds and jets. Different initial conditions, such as the tori of \citet{chakrabarti_1985_thickdisk} and \citet{penna_2013_torussolution}, may lead to distinct transport and mixing properties. Larger and more extended tori, like those explored by \citet{cho_2025_largelong}, multizone or ``cyclic zoom'' methods \citep{guo_2023_towardhorizonscale,cho_2023_scalesbondi,cho_2024_multizone,guo_2025_cycliczoom,cho_2025_multizone}, or alternative boundary conditions either from more realistic environments \citep{ressler_2018_stellarwinds,ressler_2020_madstellarwinds,hopkins_2024_fire} or imposed structure \citep{olivares_2023_transonic}, would enable studies of inflow and mixing over longer timescales and reduce sensitivity to the finite-mass reservoir. Such setups would also make it possible to probe mixing at larger radii, where the characteristic dynamical times approach thousands of $GM/c^3$ and where simulations require considerably longer to move beyond their initial transients.

Higher resolution, both the GRMHD mesh and the number of tracer particles, would allow more stringent convergence tests and may better capture the small-scale structure of turbulent transport. Incorporating additional physics, such as radiation, resistivity, or generalized nonthermal particle populations, would open the door to modeling radiatively efficient disks and exploring multi-fluid dynamics like those required to study mixed electron–ion–pair plasmas. These extensions would not only refine our physical understanding of mass transport in black hole accretion flows but also strengthen the link to observational modeling and subgrid prescriptions.

\begin{acknowledgements}
The authors thank Charles Gammie and Eliot Quataert for their useful comments and helpful discussions.
G.N.W.~was supported by the Taplin Fellowship and the Princeton Gravity Initiative.
L.M.~gratefully acknowledges support from a NASA Hubble Fellowship Program, Einstein Fellowship under award number HST-HF2-51539.001-A and NSF AST-2407810.\;
Some of the GRMHD simulations presented in this paper were run on the {\tt Ocelote} HPC resources supported by the University of Arizona TRIF, UITS, and Research, Innovation, and Impact (RII) and maintained by the UArizona Research Technologies department. Other simulations were run using the Delta advanced computing and data resource, which is supported by the U.S.~National Science Foundation (award OAC 2005572) and the State of Illinois through allocation PHY240218 from the ACCESS program, which is supported by U.S.~National Science Foundation grants \#2138259, \#2138286, \#2138307, \#2137603, and \#2138296.
\end{acknowledgements}

\appendix

\section{Simulation grid comparison}
\label{sec:appendix_grid_comparison}

\begin{figure*}
\begin{centering}
\includegraphics[width=0.85\linewidth]{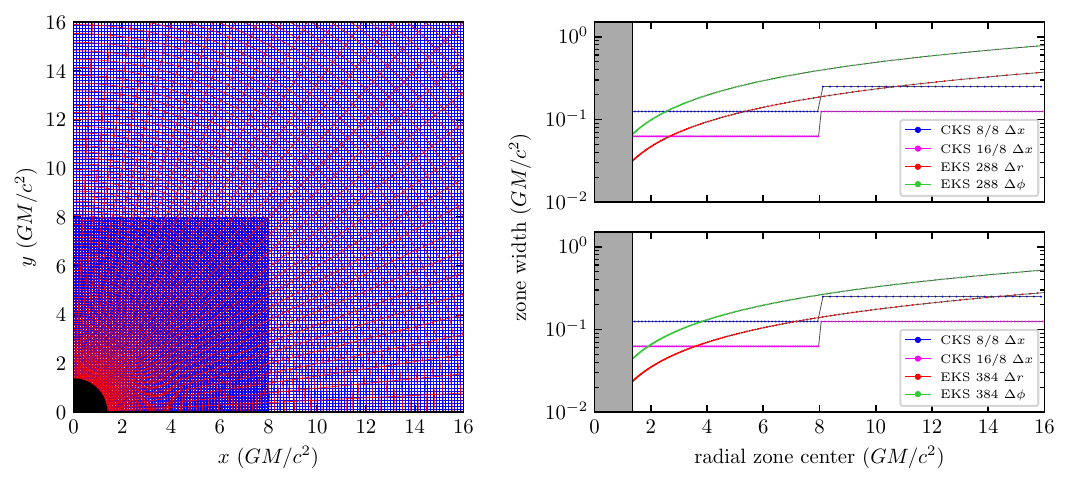}
\caption{
Comparison of grid zone size between Cartesian Kerr-Schild coordinates (CKS; blue) and Exponential spherical Kerr-Schild coordinates (EKS; red). Left: Grid zone sizes in the midplane showing the $16/8$ CKS simulation grid and the EKS $(n_r, n_\theta, n_\phi = 288,128,128)$ simulation grid often used in spherical-grid-based codes \citep{eht_m87_5,eht_m87_8,eht_sgra_5,eht_m87_9,eht_sgra_8,wong_2022_patoka,dhruv_2025_surveyv3}. Right: Comparison between zone widths for CKS $8/8$ and $16/8$ grids against reference EKS grid resolutions $288, 128, 128$ (top) and $384, 192, 192$ (bottom). The EKS grids have non-zero aspect ratios, so zone widths in both the $\hat{r}$ and $\hat{\phi}$ directions are included as separate lines. Circles mark the location of each grid zone.
}
\label{fig:grids}
\end{centering}
\end{figure*}

\begin{figure*}
\begin{centering}
\includegraphics[width=\linewidth]{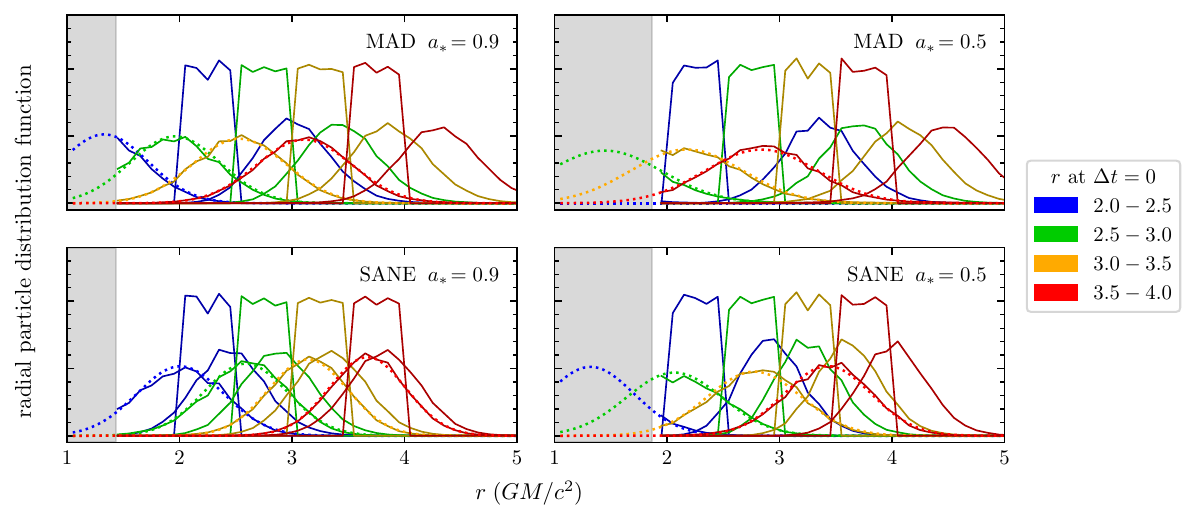}
\caption{
Example of how we estimate the mean and spread from truncated particle distributions near the event horizon. Panels show the evolution of four different particle distributions (colors) over time (lines of the same color) across four different simulations (panels). The distributions are color coded based on which radial shell they occupy at zero-time-offset $\Delta{}t = 0$. Each distribution function is plotted three times: once when the particles are within the target radial shell (top-hat shape) as well as one step before that time and one step after. As time advances, the distributions advect inward and broaden, and the left tails truncate as particles cross the event horizon. To correct for biases due to the truncation, we fit the untruncated portion of each distribution with a rescaled copy of the distribution one step before the particles are in the target radial shell (see main text). 
Solid lines show measured histograms; dotted lines show the reconstructed fits. This procedure recovers the centroid and width of the distributions despite the incomplete particle position data.
}
\label{fig:fit_spread}
\end{centering}
\end{figure*}

A practical difference between \athenak and other GRMHD codes such as HARM \citep{gammie_2003_harm} and its GPU-enabled successor Kharma \citep{prather_2024_kharma} lies in the choice of coordinate system. \athenak uses Cartesian Kerr-Schild (CKS) coordinates, while HARM and Kharma adopt Exponential Spherical Kerr-Schild (EKS) coordinates. Both frameworks rely on logically Cartesian meshes and are related to Kerr-Schild coordinates, ensuring horizon-penetrating coverage of the black hole spacetime.

Cartesian grids, like the one used in \athenak, avoid the coordinate singularities of spherical systems at the poles and origin. This eliminates the complications of implementing transparent transmissive polar boundary conditions and circumvents the severe timestep restrictions imposed by the Courant condition on the tiny near-polar zones of spherical grids. By contrast, EKS grids leverage spherical symmetry by discretizing the domain in exponentially stretched radial coordinates, uniform in azimuth, and with angular resolution concentrated toward the equatorial plane. This design provides high resolution near the black hole, where dynamical timescales are shortest, while progressively coarsening the grid at large radii, where demands are weaker. Codes such as HARM and Kharma typically employ EKS grids with logarithmic radial mapping (e.g., $n_r = 288$, $n_\theta = 128$, $n_\phi = 128$). The trade-off is that cell aspect ratios can become highly nonuniform, especially off the midplane, requiring extra care in reconstruction and flux calculations near coordinate singularities. As emphasized by \citet{stone_2024_athenak}, differences between CKS and curvilinear discretizations appear only at the level of truncation error, which vanishes in well-resolved regimes, although non-uniform truncation errors at the poles can imprint features on the flow.

Figure~\ref{fig:grids} provides a direct comparison of zone sizes between CKS and EKS grids. The left panel shows the absolute zone sizes in the midplane for a representative CKS $16/8$ grid and a conventional EKS $288 \times 128 \times 128$ grid. The right panel compares the zone widths of CKS grids ($8/8$, $16/8$) with EKS grids of different angular resolutions. While EKS achieves finer radial resolution per zone extremely close to the horizon, CKS grids offer more uniform spacing and zone shapes and typically provide more resolution at larger radii. CKS zones also remain cubical across the entire domain, whereas EKS zones are anisotropic, with differing effective resolutions in radial and azimuthal directions.

As a comparison between different simulation grids, the \athenak 16/8 CKS resolution has approximately $1.2 \times 10^8$ zones and requires about $2.1 \times 10^6$ cycles to reach $t = 40,\!000\ GM/c^3$ for a total of $\approx 2.6 \times 10^{14}$ zone updates. In contrast, the Kharma $384\times192\times192$ EKS grid has only approximately $1.4 \times 10^7$ zones but requires $1.2 \times 10^7$ cycles to reach $40,\!000\ GM/c^3$ for a total of $\approx 1.72 \times 10^{12}$ zone updates. Thus, for a comparable computational cost and total simulation time, CKS grids have a longer timestep and significantly more zones, which enables increased resolution at larger radii at the cost of slightly worse resolution very close to the event horizon.

\section{Fitting the distribution close to the event horizon}
\label{sec:appendix_fitting}

The tracer particle distributions measured in the output data truncate near the black hole since particles stopped being tracked shortly after they pass through the event horizon. Some of our analysis is based on characterizing the evolution of the particle distribution function, however, and naively measuring the statistics of the truncated distributions would artificially increase our estimate of the distribution centroid as well as lead to an underestimate of spreading. We therefore implement a simple method to estimate the part of the distribution that is within the event horizon.

By definition, the particles passing through the radial shell (at a time offset $\Delta{}t=0$) belong to a top hat distribution. We assume that the spread is roughly symmetric about $\Delta{}t=0$ and use the shape of the distribution at one timestep before $\Delta{}t=0$ (hereafter $t_{-1}$) to infer the shape one timestep after $\Delta{}t=0$ (hereafter $t_{+1}$). The number of particles in the distribution function should not change, so we only need to fit for a new center and rescaling of width. To do this, we perform a least-squares fit of the distribution at $t_{-1}$ to the untruncated part of the distribution at $t_{+1}$ (i.e., the part of the distribution outside of the event horizon).

The results of this procedure are illustrated in Figure~\ref{fig:fit_spread} for four different simulations. Each panel shows how the distribution function for particles passing through a particular radial shell (denoted by color) changes with time. The top hat, at $t=0$, identifies which radial shell each color corresponds to. The solid lines show the true particle histograms recovered from the output data. The dotted lines show the estimates for the distributions recovered using the procedure described above. 

By systematically applying this procedure, we can extend measurements of the mean trajectory and radial spread all the way down to the black hole while minimizing biases from particle loss. To avoid including spurious results, however, we discard fits whenever excluding an additional data point shifts the fitted moments by more than $5\%$, even if the reconstructed distributions appear visually well-behaved.

\bibliography{main}

\begin{thebibliography}{}
\expandafter\ifx\csname natexlab\endcsname\relax\def\natexlab#1{#1}\fi
\providecommand{\url}[1]{\href{#1}{#1}}
\providecommand{\dodoi}[1]{doi:~\href{http://doi.org/#1}{\nolinkurl{#1}}}
\providecommand{\doeprint}[1]{\href{http://ascl.net/#1}{\nolinkurl{http://ascl.net/#1}}}
\providecommand{\doarXiv}[1]{\href{https://arxiv.org/abs/#1}{\nolinkurl{https://arxiv.org/abs/#1}}}

\bibitem[{Balbus \& Hawley(1991)}]{balbus_1991_mri}
Balbus, S.~A., \& Hawley, J.~F. 1991, The Astrophysical Journal, 376, 214,
  \dodoi{10.1086/170270}

\bibitem[{{Bisnovatyi-Kogan} \& Ruzmaikin(1974)}]{bisnovatyi_1974_madstar}
{Bisnovatyi-Kogan}, G.~S., \& Ruzmaikin, A.~A. 1974, Astrophysics and Space
  Science, 28, 45, \dodoi{10.1007/BF00642237}

\bibitem[{Carballido {et~al.}(2005)Carballido, Stone, \&
  Pringle}]{carballido_2005_localdiffusion}
Carballido, A., Stone, J.~M., \& Pringle, J.~E. 2005, Monthly Notices of the
  Royal Astronomical Society, 358, 1055,
  \dodoi{10.1111/j.1365-2966.2005.08850.x}

\bibitem[{Chakrabarti(1985)}]{chakrabarti_1985_thickdisk}
Chakrabarti, S.~K. 1985, The Astrophysical Journal, 288, 1,
  \dodoi{10.1086/162755}

\bibitem[{Chandrasekhar(1960)}]{chandrasekhar_1960_mri}
Chandrasekhar, S. 1960, Proceedings of the National Academy of Science, 46,
  253, \dodoi{10.1073/pnas.46.2.253}

\bibitem[{Cho \& Narayan(2025)}]{cho_2025_largelong}
Cho, H., \& Narayan, R. 2025, Variability in {{Black Hole Accretion}}:
  {{Dependence}} on {{Rotational}} and {{Magnetic Energy Balance}},  arXiv,
  \dodoi{10.48550/arXiv.2507.13441}

\bibitem[{Cho {et~al.}(2023)Cho, Prather, Narayan, Natarajan, Su, Ricarte, \&
  Chatterjee}]{cho_2023_scalesbondi}
Cho, H., Prather, B.~S., Narayan, R., {et~al.} 2023, The Astrophysical Journal,
  959, L22, \dodoi{10.3847/2041-8213/ad1048}

\bibitem[{Cho {et~al.}(2025)Cho, Prather, Narayan, Su, \&
  Natarajan}]{cho_2025_multizone}
Cho, H., Prather, B.~S., Narayan, R., Su, K.-Y., \& Natarajan, P. 2025,
  Bridging {{Scales}} in {{Black Hole Accretion}} and {{Feedback}}:
  {{Relativistic Jet}} Linking the {{Horizon}} to the {{Host Galaxy}},  arXiv,
  \dodoi{10.48550/arXiv.2507.17818}

\bibitem[{Cho {et~al.}(2024)Cho, Prather, Su, Narayan, \&
  Natarajan}]{cho_2024_multizone}
Cho, H., Prather, B.~S., Su, K.-Y., Narayan, R., \& Natarajan, P. 2024, The
  Astrophysical Journal, 977, 200, \dodoi{10.3847/1538-4357/ad9561}

\bibitem[{Cipolletta {et~al.}(2021)Cipolletta, Kalinani, Giangrandi,
  Giacomazzo, Ciolfi, Sala, \& Giudici}]{cipolletta_2021_spritz}
Cipolletta, F., Kalinani, J.~V., Giangrandi, E., {et~al.} 2021, Classical and
  Quantum Gravity, 38, 085021, \dodoi{10.1088/1361-6382/abebb7}

\bibitem[{Del~Zanna {et~al.}(2007)Del~Zanna, Zanotti, Bucciantini, \&
  Londrillo}]{delzanna_2007_ECHOEulerianConservative}
Del~Zanna, L., Zanotti, O., Bucciantini, N., \& Londrillo, P. 2007, Astronomy
  and Astrophysics, 473, 11, \dodoi{10.1051/0004-6361:20077093}

\bibitem[{Dhruv {et~al.}(2025)Dhruv, Prather, Wong, \&
  Gammie}]{dhruv_2025_surveyv3}
Dhruv, V., Prather, B., Wong, G.~N., \& Gammie, C.~F. 2025, The Astrophysical
  Journal Supplement Series, 277, 16, \dodoi{10.3847/1538-4365/adaea6}

\bibitem[{Evans \& Hawley(1988)}]{evans_1988_ct}
Evans, C.~R., \& Hawley, J.~F. 1988, The Astrophysical Journal, 332, 659,
  \dodoi{10.1086/166684}

\bibitem[{{Event Horizon Telescope Collaboration} {et~al.}(2019){Event Horizon
  Telescope Collaboration}, Akiyama, Alberdi, Alef, Asada, Azulay, Baczko,
  Ball, Balokovi{\'c}, Barrett, Bintley, Blackburn, Boland, Bouman, Bower,
  Bremer, Brinkerink, Brissenden, Britzen, Broderick, Broguiere, Bronzwaer,
  Byun, Carlstrom, Chael, Chan, Chatterjee, Chatterjee, Chen, Chen, Cho,
  Christian, Conway, Cordes, Crew, Cui, Davelaar, De~Laurentis, Deane, Dempsey,
  Desvignes, Dexter, Doeleman, Eatough, Falcke, Fish, Fomalont,
  {Fraga-Encinas}, Friberg, Fromm, G{\'o}mez, Galison, Gammie, Garc{\'i}a,
  Gentaz, Georgiev, Goddi, Gold, Gu, Gurwell, Hada, Hecht, Hesper, Ho, Ho,
  Honma, Huang, Huang, Hughes, Ikeda, Inoue, Issaoun, James, Jannuzi, Janssen,
  Jeter, Jiang, Johnson, Jorstad, Jung, Karami, Karuppusamy, Kawashima,
  Keating, Kettenis, Kim, Kim, Kim, Kino, Koay, Koch, Koyama, Kramer, Kramer,
  Krichbaum, Kuo, Lauer, Lee, Li, Li, Lindqvist, Liu, Liuzzo, Lo, Lobanov,
  Loinard, Lonsdale, Lu, MacDonald, Mao, Markoff, Marrone, Marscher,
  {Mart{\'i}-Vidal}, Matsushita, Matthews, Medeiros, Menten, Mizuno, Mizuno,
  Moran, Moriyama, Moscibrodzka, Mu{\"l}ler, Nagai, Nagar, Nakamura, Narayan,
  Narayanan, Natarajan, Neri, Ni, Noutsos, Okino, Olivares, Oyama, {\"O}zel,
  Palumbo, Patel, Pen, Pesce, Pi{\'e}tu, Plambeck, PopStefanija, Porth,
  Prather, {Preciado-L{\'o}pez}, Psaltis, Pu, Ramakrishnan, Rao, Rawlings,
  Raymond, Rezzolla, Ripperda, Roelofs, Rogers, Ros, Rose, Roshanineshat,
  Rottmann, Roy, Ruszczyk, Ryan, Rygl, S{\'a}nchez, {S{\'a}nchez-Arguelles},
  Sasada, Savolainen, Schloerb, Schuster, Shao, Shen, Small, Sohn, SooHoo,
  Tazaki, Tiede, Tilanus, Titus, Toma, Torne, Trent, Trippe, Tsuda, {van
  Bemmel}, {van Langevelde}, {van Rossum}, Wagner, Wardle, Weintroub, Wex,
  Wharton, Wielgus, Wong, Wu, Young, Young, Younsi, Yuan, Yuan, Zensus, Zhao,
  Zhao, Zhu, Anczarski, Baganoff, Eckart, Farah, Haggard, {Meyer-Zhao},
  Michalik, Nadolski, Neilsen, Nishioka, Nowak, Pradel, Primiani, Souccar,
  Vertatschitsch, Yamaguchi, \& Zhang}]{eht_m87_5}
{Event Horizon Telescope Collaboration}, Akiyama, K., Alberdi, A., {et~al.}
  2019, The Astrophysical Journal, 875, L5, \dodoi{10.3847/2041-8213/ab0f43}

\bibitem[{{Event Horizon Telescope Collaboration} {et~al.}(2021){Event Horizon
  Telescope Collaboration}, Akiyama, Algaba, Alberdi, Alef, Anantua, Asada,
  Azulay, Baczko, Ball, Balokovi{\'c}, Barrett, Benson, Bintley, Blackburn,
  Blundell, Boland, Bouman, Bower, Boyce, Bremer, Brinkerink, Brissenden,
  Britzen, Broderick, Broguiere, Bronzwaer, Byun, Carlstrom, Chael, Chan,
  Chatterjee, Chatterjee, Chen, Chen, Chesler, Cho, Christian, Conway, Cordes,
  Crawford, Crew, {Cruz-Osorio}, Cui, Davelaar, De~Laurentis, Deane, Dempsey,
  Desvignes, Dexter, Doeleman, Eatough, Falcke, Farah, Fish, Fomalont, Ford,
  {Fraga-Encinas}, Friberg, Fromm, Fuentes, Galison, Gammie, Garc{\'i}a,
  Gelles, Gentaz, Georgiev, Goddi, Gold, G{\'o}mez, {G{\'o}mez-Ruiz}, Gu,
  Gurwell, Hada, Haggard, Hecht, Hesper, Himwich, Ho, Ho, Honma, Huang, Huang,
  Hughes, Ikeda, Inoue, Issaoun, James, Jannuzi, Janssen, Jeter, Jiang,
  {Jimenez-Rosales}, Johnson, Jorstad, Jung, Karami, Karuppusamy, Kawashima,
  Keating, Kettenis, Kim, Kim, Kim, Kim, Kino, Koay, Kofuji, Koch, Koyama,
  Kramer, Kramer, Krichbaum, Kuo, Lauer, Lee, Levis, Li, Li, Lindqvist, Lico,
  Lindahl, Liu, Liu, Liuzzo, Lo, Lobanov, Loinard, Lonsdale, Lu, MacDonald,
  Mao, Marchili, Markoff, Marrone, Marscher, {Mart{\'i}-Vidal}, Matsushita,
  Matthews, Medeiros, Menten, Mizuno, Mizuno, Moran, Moriyama, Moscibrodzka,
  M{\"u}ller, Musoke, Mus~Mej{\'i}as, Michalik, Nadolski, Nagai, Nagar,
  Nakamura, Narayan, Narayanan, Natarajan, Nathanail, Neilsen, Neri, Ni,
  Noutsos, Nowak, Okino, Olivares, {Ortiz-Le{\'o}n}, Oyama, {\"O}zel, Palumbo,
  Park, Patel, Pen, Pesce, Pi{\'e}tu, Plambeck, PopStefanija, Porth, P{\"o}tzl,
  Prather, {Preciado-L{\'o}pez}, Psaltis, Pu, Ramakrishnan, Rao, Rawlings,
  Raymond, Rezzolla, Ricarte, Ripperda, Roelofs, Rogers, Ros, Rose,
  Roshanineshat, Rottmann, Roy, Ruszczyk, Rygl, S{\'a}nchez,
  {S{\'a}nchez-Arguelles}, Sasada, Savolainen, Schloerb, Schuster, Shao, Shen,
  Small, Sohn, SooHoo, Sun, Tazaki, Tetarenko, Tiede, Tilanus, Titus, Toma,
  Torne, Trent, Traianou, Trippe, {van Bemmel}, {van Langevelde}, {van Rossum},
  Wagner, {Ward-Thompson}, Wardle, Weintroub, Wex, Wharton, Wielgus, Wong, Wu,
  Yoon, Young, Young, Younsi, Yuan, Yuan, Zensus, Zhao, \& Zhao}]{eht_m87_8}
{Event Horizon Telescope Collaboration}, Akiyama, K., Algaba, J.~C., {et~al.}
  2021, The Astrophysical Journal, 910, L13, \dodoi{10.3847/2041-8213/abe4de}

\bibitem[{{Event Horizon Telescope Collaboration} {et~al.}(2022){Event Horizon
  Telescope Collaboration}, Akiyama, Alberdi, Alef, Algaba, Anantua, Asada,
  Azulay, Bach, Baczko, Ball, Balokovi{\'c}, Barrett, Baub{\"o}ck, Benson,
  Bintley, Blackburn, Blundell, Bouman, Bower, Boyce, Bremer, Brinkerink,
  Brissenden, Britzen, Broderick, Broguiere, Bronzwaer, Bustamante, Byun,
  Carlstrom, Ceccobello, Chael, Chan, Chatterjee, Chatterjee, Chen, Chen,
  Cheng, Cho, Christian, Conroy, Conway, Cordes, Crawford, Crew, {Cruz-Osorio},
  Cui, Davelaar, De~Laurentis, Deane, Dempsey, Desvignes, Dexter, Dhruv,
  Doeleman, Dougal, Dzib, Eatough, Emami, Falcke, Farah, Fish, Fomalont, Ford,
  {Fraga-Encinas}, Freeman, Friberg, Fromm, Fuentes, Galison, Gammie,
  Garc{\'i}a, Gentaz, Georgiev, Goddi, Gold, {G{\'o}mez-Ruiz}, G{\'o}mez, Gu,
  Gurwell, Hada, Haggard, Haworth, Hecht, Hesper, Heumann, Ho, Ho, Honma,
  Huang, Huang, Hughes, Ikeda, Violette~Impellizzeri, Inoue, Issaoun, James,
  Jannuzi, Janssen, Jeter, Jiang, {Jim{\'e}nez-Rosales}, Johnson, Jorstad,
  Joshi, Jung, Karami, Karuppusamy, Kawashima, Keating, Kettenis, Kim, Kim,
  Kim, Kim, Kino, Koay, Kocherlakota, Kofuji, Koch, Koyama, Kramer, Kramer,
  Krichbaum, Kuo, La~Bella, Lauer, Lee, Lee, Leung, Levis, Li, Lico, Lindahl,
  Lindqvist, Lisakov, Liu, Liu, Liuzzo, Lo, Lobanov, Loinard, Lonsdale, Lu,
  Mao, Marchili, Markoff, Marrone, Marscher, {Mart{\'i}-Vidal}, Matsushita,
  Matthews, Medeiros, Menten, Michalik, Mizuno, Mizuno, Moran, Moriyama,
  Moscibrodzka, M{\"u}ller, Mus, Musoke, Myserlis, Nadolski, Nagai, Nagar,
  Nakamura, Narayan, Narayanan, Natarajan, Nathanail, Navarro~Fuentes, Neilsen,
  Neri, Ni, Noutsos, Nowak, Oh, Okino, Olivares, {Ortiz-Le{\'o}n}, Oyama,
  {\"O}zel, Palumbo, Filippos~Paraschos, Park, Parsons, Patel, Pen, Pesce,
  Pi{\'e}tu, Plambeck, PopStefanija, Porth, P{\"o}tzl, Prather,
  {Preciado-L{\'o}pez}, Psaltis, Pu, Ramakrishnan, Rao, Rawlings, Raymond,
  Rezzolla, Ricarte, Ripperda, Roelofs, Rogers, Ros, {Romero-Ca{\~n}izales},
  Roshanineshat, Rottmann, Roy, Ruiz, Ruszczyk, Rygl, S{\'a}nchez,
  {S{\'a}nchez-Arg{\"u}elles}, {S{\'a}nchez-Portal}, Sasada, Satapathy,
  Savolainen, Schloerb, Schonfeld, Schuster, Shao, Shen, Small, Sohn, SooHoo,
  Souccar, Sun, Tazaki, Tetarenko, Tiede, Tilanus, Titus, Torne, Traianou,
  Trent, Trippe, Turk, {van Bemmel}, {van Langevelde}, {van Rossum}, Vos,
  Wagner, {Ward-Thompson}, Wardle, Weintroub, Wex, Wharton, Wielgus, Wiik,
  Witzel, Wondrak, Wong, Wu, Yamaguchi, Yoon, Young, Young, Younsi, Yuan, Yuan,
  Zensus, Zhang, Zhao, Zhao, Chan, Qiu, Ressler, \& White}]{eht_sgra_5}
{Event Horizon Telescope Collaboration}, Akiyama, K., Alberdi, A., {et~al.}
  2022, The Astrophysical Journal, 930, L16, \dodoi{10.3847/2041-8213/ac6672}

\bibitem[{{Event Horizon Telescope Collaboration} {et~al.}(2023){Event Horizon
  Telescope Collaboration}, Akiyama, Alberdi, Alef, Algaba, Anantua, Asada,
  Azulay, Bach, Baczko, Ball, Balokovi{\'c}, Barrett, Baub{\"o}ck, Benson,
  Bintley, Blackburn, Blundell, Bouman, Bower, Boyce, Bremer, Brinkerink,
  Brissenden, Britzen, Broderick, Broguiere, Bronzwaer, Bustamante, Byun,
  Carlstrom, Ceccobello, Chael, Chan, Chang, Chatterjee, Chatterjee, Chen,
  Chen, Cheng, Cho, Christian, Conroy, Conway, Cordes, Crawford, Crew,
  {Cruz-Osorio}, Cui, Dahale, Davelaar, De~Laurentis, Deane, Dempsey,
  Desvignes, Dexter, Dhruv, Doeleman, Dougal, Dzib, Eatough, Emami, Falcke,
  Farah, Fish, Fomalont, Ford, Foschi, {Fraga-Encinas}, Freeman, Friberg,
  Fromm, Fuentes, Galison, Gammie, Garc{\'i}a, Gentaz, Georgiev, Goddi, Gold,
  {G{\'o}mez-Ruiz}, G{\'o}mez, Gu, Gurwell, Hada, Haggard, Haworth, Hecht,
  Hesper, Heumann, Ho, Ho, Honma, Huang, Huang, Hughes, Ikeda, Impellizzeri,
  Inoue, Issaoun, James, Jannuzi, Janssen, Jeter, Jiang, {Jim{\'e}nez-Rosales},
  Johnson, Jorstad, Joshi, Jung, Karami, Karuppusamy, Kawashima, Keating,
  Kettenis, Kim, Kim, Kim, Kim, Kino, Koay, Kocherlakota, Kofuji, Koch, Koyama,
  Kramer, Kramer, Kramer, Krichbaum, Kuo, La~Bella, Lauer, Lee, Lee, Leung,
  Levis, Li, Lico, Lindahl, Lindqvist, Lisakov, Liu, Liu, Liuzzo, Lo, Lobanov,
  Loinard, Lonsdale, Lowitz, Lu, MacDonald, Mao, Marchili, Markoff, Marrone,
  Marscher, {Mart{\'i}-Vidal}, Matsushita, Matthews, Medeiros, Menten,
  Michalik, Mizuno, Mizuno, Moran, Moriyama, Moscibrodzka, Mulaudzi,
  M{\"u}ller, M{\"u}ller, Mus, Musoke, Myserlis, Nadolski, Nagai, Nagar,
  Nakamura, Narayan, Narayanan, Natarajan, Nathanail, Fuentes, Neilsen, Neri,
  Ni, Noutsos, Nowak, Oh, Okino, Olivares, {Ortiz-Le{\'o}n}, Oyama, {\"O}zel,
  Palumbo, Paraschos, Park, Parsons, Patel, Pen, Pesce, Pi{\'e}tu, Plambeck,
  PopStefanija, Porth, P{\"o}tzl, Prather, {Preciado-L{\'o}pez}, Psaltis, Pu,
  Ramakrishnan, Rao, Rawlings, Raymond, Rezzolla, Ricarte, Ripperda, Roelofs,
  Rogers, {Romero-Ca{\~n}izales}, Ros, Roshanineshat, Rottmann, Roy, Ruiz,
  Ruszczyk, Rygl, S{\'a}nchez, {S{\'a}nchez-Arg{\"u}elles},
  {S{\'a}nchez-Portal}, Sasada, Satapathy, Savolainen, Schloerb, Schonfeld,
  Schuster, Shao, Shen, Small, Sohn, SooHoo, Sosapanta~Salas, Souccar, Sun,
  Tazaki, Tetarenko, Tiede, Tilanus, Titus, Torne, Toscano, Traianou, Trent,
  Trippe, Turk, {van Bemmel}, {van Langevelde}, {van Rossum}, Vos, Wagner,
  {Ward-Thompson}, Wardle, Washington, Weintroub, Wharton, Wielgus, Wiik,
  Witzel, Wondrak, Wong, Wu, Yadlapalli, Yamaguchi, Yfantis, Yoon, Young,
  Young, Younsi, Yu, Yuan, Yuan, Zensus, Zhang, Zhao, \& Zhao}]{eht_m87_9}
---. 2023, The Astrophysical Journal, 957, L20,
  \dodoi{10.3847/2041-8213/acff70}

\bibitem[{{Event Horizon Telescope Collaboration} {et~al.}(2024){Event Horizon
  Telescope Collaboration}, Akiyama, Alberdi, Alef, Algaba, Anantua, Asada,
  Azulay, Bach, Baczko, Ball, Balokovi{\'c}, Bandyopadhyay, Barrett,
  Baub{\"o}ck, Benson, Bintley, Blackburn, Blundell, Bouman, Bower, Boyce,
  Bremer, Brinkerink, Brissenden, Britzen, Broderick, Broguiere, Bronzwaer,
  Bustamante, Byun, Carlstrom, Ceccobello, Chael, Chan, Chang, Chatterjee,
  Chatterjee, Chen, Chen, Cheng, Cho, Christian, Conroy, Conway, Cordes,
  Crawford, Crew, {Cruz-Osorio}, Cui, Dahale, Davelaar, De~Laurentis, Deane,
  Dempsey, Desvignes, Dexter, Dhruv, Dihingia, Doeleman, Dougall, Dzib,
  Eatough, Emami, Falcke, Farah, Fish, Fomalont, Ford, Foschi, {Fraga-Encinas},
  Freeman, Friberg, Fromm, Fuentes, Galison, Gammie, Garc{\'i}a, Gentaz,
  Georgiev, Goddi, Gold, {G{\'o}mez-Ruiz}, G{\'o}mez, Gu, Gurwell, Hada,
  Haggard, Haworth, Hecht, Hesper, Heumann, Ho, Ho, Honma, Huang, Huang,
  Hughes, Ikeda, Impellizzeri, Inoue, Issaoun, James, Jannuzi, Janssen, Jeter,
  Jiang, {Jim{\'e}nez-Rosales}, Johnson, Jorstad, Joshi, Jung, Karami,
  Karuppusamy, Kawashima, Keating, Kettenis, Kim, Kim, Kim, Kim, Kino, Koay,
  Kocherlakota, Kofuji, Koch, Koyama, Kramer, Kramer, Kramer, Krichbaum, Kuo,
  La~Bella, Lauer, Lee, Lee, Leung, Levis, Li, Lico, Lindahl, Lindqvist,
  Lisakov, Liu, Liu, Liuzzo, Lo, Lobanov, Loinard, Lonsdale, Lowitz, Lu,
  MacDonald, Mao, Marchili, Markoff, Marrone, Marscher, {Mart{\'i}-Vidal},
  Matsushita, Matthews, Medeiros, Menten, Michalik, Mizuno, Mizuno, Moran,
  Moriyama, Moscibrodzka, Mulaudzi, M{\"u}ller, M{\"u}ller, Mus, Musoke,
  Myserlis, Nadolski, Nagai, Nagar, Nakamura, Narayanan, Natarajan, Nathanail,
  Fuentes, Neilsen, Neri, Ni, Noutsos, Nowak, Oh, Okino, Olivares,
  {Ortiz-Le{\'o}n}, Oyama, {\"O}zel, Palumbo, Paraschos, Park, Parsons, Patel,
  Pen, Pesce, Pi{\'e}tu, Plambeck, PopStefanija, Porth, P{\"o}tzl, Prather,
  {Preciado-L{\'o}pez}, Psaltis, Pu, Ramakrishnan, Rao, Rawlings, Raymond,
  Rezzolla, Ricarte, Ripperda, Roelofs, Rogers, {Romero-Ca{\~n}izales}, Ros,
  Roshanineshat, Rottmann, Roy, Ruiz, Ruszczyk, Rygl, S{\'a}nchez,
  {S{\'a}nchez-Arg{\"u}elles}, {S{\'a}nchez-Portal}, Sasada, Satapathy,
  Savolainen, Schloerb, Schonfeld, Schuster, Shao, Shen, Small, Sohn, SooHoo,
  Sosapanta~Salas, Souccar, Stanway, Sun, Tazaki, Tetarenko, Tiede, Tilanus,
  Titus, Torne, Toscano, Traianou, Trent, Trippe, Turk, {van Bemmel}, {van
  Langevelde}, {van Rossum}, Vos, Wagner, {Ward-Thompson}, Wardle, Washington,
  Weintroub, Wharton, Wielgus, Wiik, Witzel, Wondrak, Wong, Wu, Yadlapalli,
  Yamaguchi, Yfantis, Yoon, Young, Young, Younsi, Yu, Yuan, Yuan, Zensus,
  Zhang, Zhao, Zhao, \& {Najafi-Ziyazi}}]{eht_sgra_8}
---. 2024, The Astrophysical Journal, 964, L26,
  \dodoi{10.3847/2041-8213/ad2df1}

\bibitem[{Fern{\'a}ndez {et~al.}(2015)Fern{\'a}ndez, Kasen, Metzger, \&
  Quataert}]{fernandez_2015_outflowsnsmergers}
Fern{\'a}ndez, R., Kasen, D., Metzger, B.~D., \& Quataert, E. 2015, Monthly
  Notices of the Royal Astronomical Society, 446, 750,
  \dodoi{10.1093/mnras/stu2112}

\bibitem[{Fern{\'a}ndez \& Metzger(2013)}]{fernandez_2013_delayed}
Fern{\'a}ndez, R., \& Metzger, B.~D. 2013, Monthly Notices of the Royal
  Astronomical Society, 435, 502, \dodoi{10.1093/mnras/stt1312}

\bibitem[{Fishbone \& Moncrief(1976)}]{fishbone_1976_torus}
Fishbone, L.~G., \& Moncrief, V. 1976, The Astrophysical Journal, 207, 962,
  \dodoi{10.1086/154565}

\bibitem[{Gammie(2025)}]{gammie_2025_adiabaticindex}
Gammie, C.~F. 2025, The Astrophysical Journal, 980, 193,
  \dodoi{10.3847/1538-4357/adaea3}

\bibitem[{Gammie {et~al.}(2003)Gammie, McKinney, \&
  T{\'o}th}]{gammie_2003_harm}
Gammie, C.~F., McKinney, J.~C., \& T{\'o}th, G. 2003, The Astrophysical
  Journal, 589, 444, \dodoi{10.1086/374594}

\bibitem[{Gardiner \& Stone(2005)}]{gardiner_2005_ct}
Gardiner, T.~A., \& Stone, J.~M. 2005, Journal of Computational Physics, 205,
  509, \dodoi{10.1016/j.jcp.2004.11.016}

\bibitem[{Gardiner \& Stone(2008)}]{gardiner_2008_ct}
---. 2008, Journal of Computational Physics, 227, 4123,
  \dodoi{10.1016/j.jcp.2007.12.017}

\bibitem[{Genel {et~al.}(2013)Genel, Vogelsberger, Nelson, Sijacki, Springel,
  \& Hernquist}]{genel_2013_lagrangianmc}
Genel, S., Vogelsberger, M., Nelson, D., {et~al.} 2013, Monthly Notices of the
  Royal Astronomical Society, 435, 1426, \dodoi{10.1093/mnras/stt1383}

\bibitem[{Guo {et~al.}(2023)Guo, Stone, Kim, \&
  Quataert}]{guo_2023_towardhorizonscale}
Guo, M., Stone, J.~M., Kim, C.-G., \& Quataert, E. 2023, The Astrophysical
  Journal, 946, 26, \dodoi{10.3847/1538-4357/acb81e}

\bibitem[{Guo {et~al.}(2025)Guo, Stone, Quataert, \&
  Springel}]{guo_2025_cycliczoom}
Guo, M., Stone, J.~M., Quataert, E., \& Springel, V. 2025, The Astrophysical
  Journal, 987, 202, \dodoi{10.3847/1538-4357/add1da}

\bibitem[{Hawley(2000)}]{hawley_2000_globaltori}
Hawley, J.~F. 2000, The Astrophysical Journal, 528, 462, \dodoi{10.1086/308180}

\bibitem[{Hawley {et~al.}(1995)Hawley, Gammie, \&
  Balbus}]{hawley_1995_localmri}
Hawley, J.~F., Gammie, C.~F., \& Balbus, S.~A. 1995, The Astrophysical Journal,
  440, 742, \dodoi{10.1086/175311}

\bibitem[{Ho(2008)}]{ho_2008_llagn}
Ho, L.~C. 2008, Annual Review of Astronomy and Astrophysics, 46, 475,
  \dodoi{10.1146/annurev.astro.45.051806.110546}

\bibitem[{Hopkins {et~al.}(2024)Hopkins, Grudic, Su, Wellons, {Angles-Alcazar},
  Steinwandel, Guszejnov, Murray, {Faucher-Giguere}, Quataert, \&
  Keres}]{hopkins_2024_fire}
Hopkins, P.~F., Grudic, M.~Y., Su, K.-Y., {et~al.} 2024, The Open Journal of
  Astrophysics, 7, 18, \dodoi{10.21105/astro.2309.13115}

\bibitem[{Igumenshchev {et~al.}(2003)Igumenshchev, Narayan, \&
  Abramowicz}]{igumenshchev_2003_mad}
Igumenshchev, I.~V., Narayan, R., \& Abramowicz, M.~A. 2003, The Astrophysical
  Journal, 592, 1042, \dodoi{10.1086/375769}

\bibitem[{K{\"a}pyl{\"a} {et~al.}(2009)K{\"a}pyl{\"a}, Korpi, \&
  Brandenburg}]{kapyla_2009_turbdiffusion}
K{\"a}pyl{\"a}, P.~J., Korpi, M.~J., \& Brandenburg, A. 2009, Astronomy and
  Astrophysics, 500, 633, \dodoi{10.1051/0004-6361/200811498}

\bibitem[{Lemaster \& Stone(2009)}]{lemaster_2009_fofc}
Lemaster, M.~N., \& Stone, J.~M. 2009, The Astrophysical Journal, 691, 1092,
  \dodoi{10.1088/0004-637X/691/2/1092}

\bibitem[{Liska {et~al.}(2018)Liska, Hesp, Tchekhovskoy, Ingram, {van der
  Klis}, \& Markoff}]{liska_2018_tilted}
Liska, M., Hesp, C., Tchekhovskoy, A., {et~al.} 2018, Monthly Notices of the
  Royal Astronomical Society, 474, L81, \dodoi{10.1093/mnrasl/slx174}

\bibitem[{Margalit \& Metzger(2016)}]{margalit_2016_nuclearburning}
Margalit, B., \& Metzger, B.~D. 2016, Monthly Notices of the Royal Astronomical
  Society, 461, 1154, \dodoi{10.1093/mnras/stw1410}

\bibitem[{McKinney(2006)}]{mckinney_2006_jets}
McKinney, J.~C. 2006, Monthly Notices of the Royal Astronomical Society, 368,
  1561, \dodoi{10.1111/j.1365-2966.2006.10256.x}

\bibitem[{Mignone {et~al.}(2007)Mignone, Bodo, Massaglia, Matsakos, Tesileanu,
  Zanni, \& Ferrari}]{mignone_2007_PLUTO}
Mignone, A., Bodo, G., Massaglia, S., {et~al.} 2007, The Astrophysical Journal
  Supplement Series, Volume 170, Issue 1, pp. 228-242., 170, 228,
  \dodoi{10.1086/513316}

\bibitem[{Narayan {et~al.}(2003)Narayan, Igumenshchev, \&
  Abramowicz}]{narayan_2003_mad}
Narayan, R., Igumenshchev, I.~V., \& Abramowicz, M.~A. 2003, Publications of
  the Astronomical Society of Japan, 55, L69, \dodoi{10.1093/pasj/55.6.L69}

\bibitem[{Narayan {et~al.}(2012)Narayan, S{\k a}dowski, Penna, \&
  Kulkarni}]{narayan_2012_sane}
Narayan, R., S{\k a}dowski, A., Penna, R.~F., \& Kulkarni, A.~K. 2012, Monthly
  Notices of the Royal Astronomical Society, 426, 3241,
  \dodoi{10.1111/j.1365-2966.2012.22002.x}

\bibitem[{Narayan \& Yi(1995)}]{narayan_1995_ADAF}
Narayan, R., \& Yi, I. 1995, The Astrophysical Journal, 452, 710,
  \dodoi{10.1086/176343}

\bibitem[{Novikov \& Thorne(1973)}]{novikov_1973_diskmodel}
Novikov, I.~D., \& Thorne, K.~S. 1973, Black holes (Les Astres Occlus), 343

\bibitem[{Olivares {et~al.}(2019)Olivares, Porth, Davelaar, Most, Fromm,
  Mizuno, Younsi, \& Rezzolla}]{olivares_2019_bhac}
Olivares, H., Porth, O., Davelaar, J., {et~al.} 2019, Astronomy and
  Astrophysics, 629, A61, \dodoi{10.1051/0004-6361/201935559}

\bibitem[{Olivares {et~al.}(2023)Olivares, Mo{\'s}cibrodzka, \&
  Porth}]{olivares_2023_transonic}
Olivares, H.~R., Mo{\'s}cibrodzka, M.~A., \& Porth, O. 2023, Astronomy and
  Astrophysics, 678, A141, \dodoi{10.1051/0004-6361/202346010}

\bibitem[{Penna {et~al.}(2013)Penna, Kulkarni, \&
  Narayan}]{penna_2013_torussolution}
Penna, R.~F., Kulkarni, A., \& Narayan, R. 2013, Astronomy and Astrophysics,
  559, A116, \dodoi{10.1051/0004-6361/201219666}

\bibitem[{Porth {et~al.}(2017)Porth, Olivares, Mizuno, Younsi, Rezzolla,
  Moscibrodzka, Falcke, \& Kramer}]{porth_2017_BHAC}
Porth, O., Olivares, H., Mizuno, Y., {et~al.} 2017, Computational Astrophysics
  and Cosmology, 4, 1, \dodoi{10.1186/s40668-017-0020-2}

\bibitem[{Prather(2024)}]{prather_2024_kharma}
Prather, B.~S. 2024, {{KHARMA}}: {{Flexible}}, {{Portable Performance}} for
  {{GRMHD}},  arXiv, \dodoi{10.48550/arXiv.2408.01361}

\bibitem[{Ressler {et~al.}(2018)Ressler, Quataert, \&
  Stone}]{ressler_2018_stellarwinds}
Ressler, S.~M., Quataert, E., \& Stone, J.~M. 2018, Monthly Notices of the
  Royal Astronomical Society, 478, 3544, \dodoi{10.1093/mnras/sty1146}

\bibitem[{Ressler {et~al.}(2021)Ressler, Quataert, White, \&
  Blaes}]{ressler_2021_sphericalaccretion}
Ressler, S.~M., Quataert, E., White, C.~J., \& Blaes, O. 2021, Monthly Notices
  of the Royal Astronomical Society, 504, 6076, \dodoi{10.1093/mnras/stab311}

\bibitem[{Ressler {et~al.}(2015)Ressler, Tchekhovskoy, Quataert, Chandra, \&
  Gammie}]{ressler_2015_electronthermo}
Ressler, S.~M., Tchekhovskoy, A., Quataert, E., Chandra, M., \& Gammie, C.~F.
  2015, Monthly Notices of the Royal Astronomical Society, 454, 1848,
  \dodoi{10.1093/mnras/stv2084}

\bibitem[{Ressler {et~al.}(2020)Ressler, White, Quataert, \&
  Stone}]{ressler_2020_madstellarwinds}
Ressler, S.~M., White, C.~J., Quataert, E., \& Stone, J.~M. 2020, The
  Astrophysical Journal, 896, L6, \dodoi{10.3847/2041-8213/ab9532}

\bibitem[{S{\k a}dowski {et~al.}(2013{\natexlab{a}})S{\k a}dowski, Narayan,
  Penna, \& Zhu}]{sadowski_2013_sane}
S{\k a}dowski, A., Narayan, R., Penna, R., \& Zhu, Y. 2013{\natexlab{a}},
  Monthly Notices of the Royal Astronomical Society, 436, 3856,
  \dodoi{10.1093/mnras/stt1881}

\bibitem[{S{\k a}dowski {et~al.}(2013{\natexlab{b}})S{\k a}dowski, Narayan,
  Tchekhovskoy, \& Zhu}]{sadowski_2013_koral}
S{\k a}dowski, A., Narayan, R., Tchekhovskoy, A., \& Zhu, Y.
  2013{\natexlab{b}}, Monthly Notices of the Royal Astronomical Society, 429,
  3533, \dodoi{10.1093/mnras/sts632}

\bibitem[{S{\k a}dowski {et~al.}(2017)S{\k a}dowski, Wielgus, Narayan, Abarca,
  McKinney, \& Chael}]{sadowski_2017_twotemp}
S{\k a}dowski, A., Wielgus, M., Narayan, R., {et~al.} 2017, Monthly Notices of
  the Royal Astronomical Society, 466, 705, \dodoi{10.1093/mnras/stw3116}

\bibitem[{Shakura \& Sunyaev(1973)}]{shakura_1973_disks}
Shakura, N.~I., \& Sunyaev, R.~A. 1973, Astronomy and Astrophysics, 24, 337

\bibitem[{Shankar {et~al.}(2023)Shankar, M{\"o}sta, Brandt, Haas, Schnetter, \&
  {de Graaf}}]{shankar_2023_gramx}
Shankar, S., M{\"o}sta, P., Brandt, S.~R., {et~al.} 2023, Classical and Quantum
  Gravity, 40, 205009, \dodoi{10.1088/1361-6382/acf2d9}

\bibitem[{Stone {et~al.}(1996)Stone, Hawley, Gammie, \&
  Balbus}]{stone_1996_stratified}
Stone, J.~M., Hawley, J.~F., Gammie, C.~F., \& Balbus, S.~A. 1996, The
  Astrophysical Journal, 463, 656, \dodoi{10.1086/177280}

\bibitem[{Stone {et~al.}(2024)Stone, Mullen, Fielding, Grete, Guo, Kempski,
  Most, White, \& Wong}]{stone_2024_athenak}
Stone, J.~M., Mullen, P.~D., Fielding, D., {et~al.} 2024, {{AthenaK}}: {{A
  Performance-Portable Version}} of the {{Athena}}++ {{AMR Framework}},  arXiv,
  \dodoi{10.48550/arXiv.2409.16053}

\bibitem[{Tominaga {et~al.}(2019)Tominaga, Takahashi, \&
  Inutsuka}]{tominaga_2019_dustdiffusion}
Tominaga, R.~T., Takahashi, S.~Z., \& Inutsuka, S.-i. 2019, The Astrophysical
  Journal, 881, 53, \dodoi{10.3847/1538-4357/ab25ea}

\bibitem[{Turner {et~al.}(2006)Turner, Willacy, Bryden, \&
  Yorke}]{turner_2006_mixing}
Turner, N.~J., Willacy, K., Bryden, G., \& Yorke, H.~W. 2006, The Astrophysical
  Journal, 639, 1218, \dodoi{10.1086/499486}

\bibitem[{Velikhov(1959)}]{velikhov_1959_mri}
Velikhov, E.~P. 1959, Soviet Journal of Experimental and Theoretical Physics,
  9, 995

\bibitem[{White {et~al.}(2016)White, Stone, \& Gammie}]{white_2016_athenapp}
White, C.~J., Stone, J.~M., \& Gammie, C.~F. 2016, The Astrophysical Journal
  Supplement Series, 225, 22, \dodoi{10.3847/0067-0049/225/2/22}

\bibitem[{Wong {et~al.}(2021)Wong, Du, Prather, \& Gammie}]{wong_2021_jetdisk}
Wong, G.~N., Du, Y., Prather, B.~S., \& Gammie, C.~F. 2021, The Astrophysical
  Journal, 914, 55, \dodoi{10.3847/1538-4357/abf8b8}

\bibitem[{Wong {et~al.}(2022)Wong, Prather, Dhruv, Ryan, Mo{\'s}cibrodzka,
  Chan, Joshi, Yarza, Ricarte, Shiokawa, Dolence, Noble, McKinney, \&
  Gammie}]{wong_2022_patoka}
Wong, G.~N., Prather, B.~S., Dhruv, V., {et~al.} 2022, The Astrophysical
  Journal Supplement Series, 259, 64, \dodoi{10.3847/1538-4365/ac582e}

\bibitem[{Yuan {et~al.}(2003)Yuan, Quataert, \& Narayan}]{yuan_2003_nonthermal}
Yuan, F., Quataert, E., \& Narayan, R. 2003, The Astrophysical Journal, 598,
  301, \dodoi{10.1086/378716}

\end{thebibliography}
\bibliographystyle{aasjournal}

\end{document}